\documentclass[lettersize,journal]{IEEEtran}

\usepackage{amsmath,amsfonts}
\usepackage{algorithmic}
\usepackage{algorithm}
\usepackage{array}
\usepackage[caption=false,font=normalsize,labelfont=sf,textfont=sf]{subfig}
\usepackage{textcomp}
\usepackage{stfloats}
\usepackage{url}
\usepackage{verbatim}
\usepackage{graphicx}

\usepackage{cite}
\usepackage{xcolor,colortbl}

\definecolor{Gray1}{gray}{0.85}
\definecolor{Gray2}{gray}{0.93}
\definecolor{LightCyan}{rgb}{0.88,1,1}
\newcolumntype{a}{>{\columncolor{Gray}}c}
\RequirePackage{booktabs}%

\begin{document}

\title{Power-Swing Blocking and Out-of-Step Tripping Scheme Using the Swing Center Voltage in Systems With Inverter-Based Resources}

\author{Henrik Johansson,~\IEEEmembership{Graduate Student Member,~IEEE}, Qianli Xing,~\IEEEmembership{Graduate Student Member,~IEEE}, Jaekeun Lee,~\IEEEmembership{Student Member,~IEEE}, Ki-Hyun Kim,~\IEEEmembership{Graduate Student Member,~IEEE}, Nathaniel Taylor,~\IEEEmembership{Senior Member,~IEEE}, and Xiongfei Wang,~\IEEEmembership{Fellow,~IEEE}


}

\maketitle

\begin{abstract}
Misoperation of power-swing blocking (PSB) and out-of-step tripping (OST) schemes in distance relays is a critical concern in power systems with increasing penetration levels of inverter-based resources (IBRs). Recently, several papers have studied the impact of IBRs on conventional dual-blinder PSB/OST schemes. They highlight interoperability issues concerning fast power swings and unconventional impedance trajectories, and propose revised relay settings in attempts to deal with the problems. However, the blinder-based PSB methodology suffers from being dependent on the speed of the apparent impedance just as it crosses between the blinders. It is therefore important to consider alternative methods. This paper thoroughly analyzes the characteristics of the swing center voltage (SCV) and the impacts of grid-forming (GFM) and grid-following (GFL) inverters on its local estimation. Then, an SCV-based PSB/OST scheme is proposed, where the main PSB criterion can be tuned from simple system parameters. This is made possible by a separate PSB activation function. The OST scheme uses only one setting and is motivated from the analytical analysis of the SCV with IBRs. Extensive PSCAD/EMTDC simulations verify the performance of the proposed SCV-based PSB/OST scheme and how it overcomes limitations of blinder-based schemes with GFM and GFL inverters.

\end{abstract}


\section{Introduction}

\IEEEPARstart{D}{istance} relays continuously measure the apparent impedance of the six fault loops to estimate the positive-sequence line impedance (distance) to faults occurring on the covered transmission lines. During power swings, the measured impedances may undesirably enter inside the distance trip zones, and the power-swing blocking (PSB) function should block distance relays from tripping in such cases \cite{PSB-Tutorial-Normann}. If loss of synchronism occurs, the out-of-step tripping (OST) function should detect the unstable power swing in order to execute a controlled separation of the system at predetermined generation-load balance locations to avoid equipment damage and prevent a widespread power outage through cascaded contingency tripping. Today, dual-blinder impedance-based schemes are commonly used for PSB and OST in conventional distance relays \cite{PSB-IBRs-PartII, PSB-impact-of-wind, REL670-application-manual, D60-instruction-manual-GE, SEL-421-4, PSB-Bogdan-Zcont}. These schemes assume that the measured apparent positive-sequence impedance $\bar{Z}_1$ quickly crosses the blinder zones during faults but significantly slower during power swings.

\IEEEpubidadjcol

Misoperations of conventional PSB schemes in transmission line distance relays have been prominent causes for many of the latest blackouts around the world \cite{PSB-Naidu-DPSP2024}. For example, a power outage in Mexico occurred when their transmission grid operated with a high share of renewable energy in the North regions while having a high power flow to the South regions \cite{PSB-CFE-SCV_vs_Blinders}. A power swing occurred following the loss of substantial transmission capacity due to a line fault. The implemented blinder-based PSB scheme in distance relays at other lines between the North and South regions was unable to detect the power swing and block their distance zones, resulting in unwanted tripping. Another case is the 2023 blackout event in Brazil. Misoperations of PSB and OST schemes in an area with a high penetration of renewable energy have been identified as a primary cause of this blackout \cite{PSB-Yongxin-GFL}.

Due to the continuous increase of inverter-based resources (IBRs) in modern power systems (e.g. wind and solar power), as well as their prevailing interoperability issues with installed PSB/OST schemes, several recent papers have studied the impact of both grid-following (GFL) and grid-forming (GFM) inverters on the conventional dual-blinder PSB/OST methodology \cite{PSB-impact-of-wind, PSB-IBRs-PartII, PSB-GFM-VI, PSB-Yongxin-GFL, PSB-Yongxin-GFM, PSB-GFM-Protection-and-Dynamics}. These papers conclude that the PSB and OST settings should be revised, mainly concerning faster and different $\bar{Z}_1$ trajectories compared with synchronous generators (SGs). This includes adjusting the PSB pickup time, OST logic and blinder zones. However, faster PSB assertion can cause unwanted blocking of distance relays during actual faults and compromise the system transient stability. Therefore, distance relay manufacturers tend to advise against using a pickup threshold lower than 1.5 fundamental periods \cite{PSB-IBRs-PartII}, at least for the first PSB activation. For example, in \cite{REL670-application-manual}, the PSB pickup times for the initial and subsequent power swing cycles in a $50$~Hz power system are calculated to $30$ and $10$~ms, respectively. On the other hand, Comisión Federal de Electricidad (CFE) in Mexico used a power swing detection time of $8.3$~ms ($0.5$ fundamental periods at $60~$Hz), yet even this turned out to be too slow with IBRs; the impedance crossed the outer and middle blinders within only $4.3$~ms \cite{PSB-CFE-SCV_vs_Blinders}.

It is seen that the traditional dual-blinder PSB/OST methodology has limitations concerning IBRs. Hence, it is important to investigate alternative ways to realize the PSB and OST functionalities in distance relays near IBRs. This paper analyzes the characteristics of the swing center voltage (SCV) and the impact of IBRs on its local estimation in detail. Then, an SCV-based PSB/OST scheme is proposed. Validated with extensive PSCAD/EMTDC simulations in two- and three-source systems, it operates correctly with GFM and GFL inverters when traditional dual-blinder schemes fail, and during the initial swing cycle when blinder schemes cannot if $\bar{Z}_1$ is already inside the blinders from a fault and immediately enters the distance zones---an issue often overlooked in the current literature \cite{PSB-impact-of-wind, PSB-IBRs-PartII, PSB-GFM-VI, PSB-Yongxin-GFL, PSB-Yongxin-GFM, PSB-GFM-Protection-and-Dynamics}. Situations like these can occur with SGs due to unstable or marginally stable power swings \cite{PSB-Zero-setting, REL670-application-manual}, but IBRs can increase the risk \cite{HJ-PSB-IQ2}.

Following this introduction, Section~\ref{sec:benchmark-performance-eval} describes the system used for analysis and benchmark PSCAD simulations. Section~\ref{sec:SCV-characteristics} explains the characteristics of the SCV in detail. Section~\ref{sec:Impact-IBRs-SCV} analyzes the impacts of GFM and GFL inverters on the SCV estimated by distance relays. Section~\ref{sec:Proposed-SCV-scheme} proposes the SCV-based PSB/OST scheme. Section~\ref{sec:benchmark-performance-evaluation} validates the analytical analysis and performance of the proposed SCV-based PSB/OST scheme.  Section~\ref{sec:conclusion} concludes the findings.

\section{Benchmark System Description}
\label{sec:benchmark-performance-eval}
\begin{figure}[h]
    \centering
    \includegraphics[width=1\linewidth]{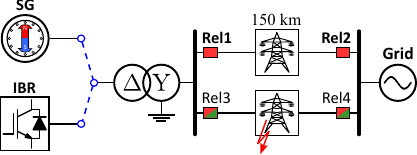}
    \caption{Benchmark two-source parallel line system to generate power swings.}
    \label{fig:two-source-parallel-ohl}
\end{figure}

The parallel-line system shown in Fig.~\ref{fig:two-source-parallel-ohl} is used for analytical analysis of the SCV with different sources and benchmark performance evaluation of the proposed SCV-based PSB/OST scheme. The sending-end source (left) is either an SG ($H=3$~s) or an IBR, and the remote-end source (right) is the utility grid with a source-to-line impedance ratio (SIR) of either $0.3$, $0.6$ or $0.9$ for each line. The analyzed quadrilateral distance relays are ``Rel1'' and ``Rel2'', which have a reach of $80\%$, $135\%$ and $200\%$ of the total line length for their zone 1, 2 and 3, respectively. Zone 1, 2 and 3 trip within 5, 450 and 1000 ms, respectively. The resistive reach extensions have been tuned to accurately detect faults when an SG is connected at the sending end (see \cite{HJ-GFM-distance} for more details). To cause power swings, balanced faults are applied along the lower line and cleared after a certain time. Both lines are modelled using the frequency-dependent phase model available in PSCAD. The system voltage, frequency and nominal power are 220~kV, 50~Hz and 300~MVA, respectively. The parallel line is 150~km and the simulated balanced fault-to-swing cases with either an SG or IBR in Fig.~\ref{fig:two-source-parallel-ohl} are:
\begin{itemize}
    \item Distance from Rel1 $m_\mathrm{f}\in \{10, 30, 75, 120, 140 \}$~km 
    \item Fault resistance $R_\mathrm{f} \in \{0, 1, 5, 10, 15\}$ $\Omega$
    \item Fault clearing time \newline $t_\mathrm{c} \in \{40, 60, 100, 150, 200, 300, 400, 500, 10000\}$~ms. 
\end{itemize}
The blinder zones will be shown in Section~\ref{sec:benchmark-performance-evaluation}. Further details concerning the blinder-based PSB/OST scheme can be found in \cite{HJ-PSB-IQ2}. However, it can be mentioned that if $\bar{Z}_1$ takes longer than $\Delta T_\mathrm{PSB,1}=30$~ms to cross between the outer and middle blinders, a PSB signal is issued, and for subsequent swing cycles, the PSB pickup time is $\Delta T_\mathrm{PSB,2}=10$~ms (these settings are used in an example in \cite{REL670-application-manual}). OST uses an inner blinder zone.

\subsection{Implemented GFM and GFL Inverter Control Strategies}
\begin{figure}[h]
    \centering
    \includegraphics[width=1\linewidth]{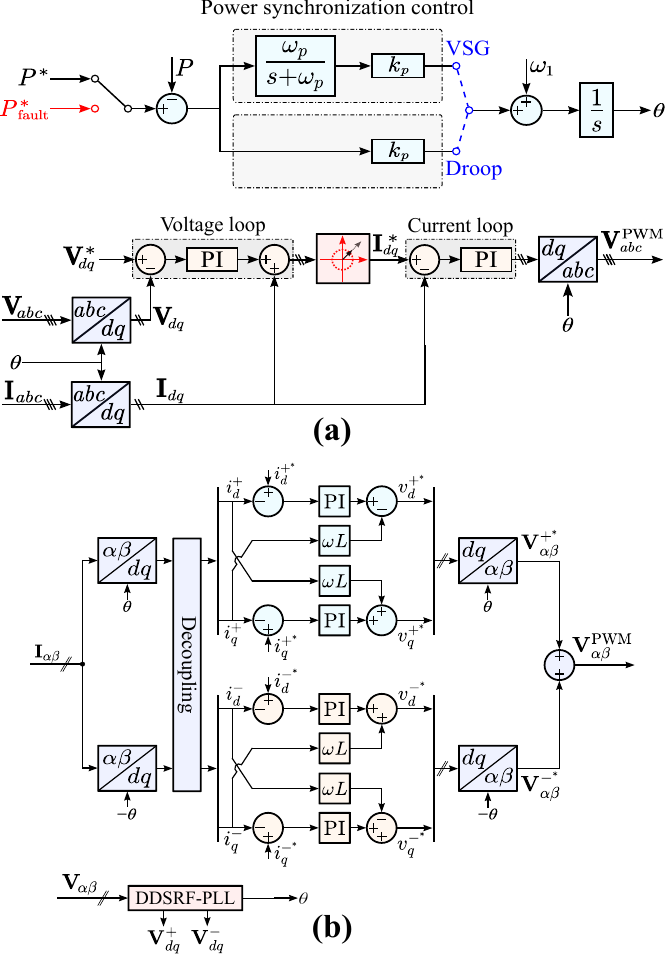}
    \caption{IBR control diagrams: \textbf{(a)} GFM inverter; \textbf{(b)} GFL inverter.}
    \label{fig:GFM-and-GFL-diagram}
\end{figure}

The GFM inverter control diagram is shown in Fig.~\ref{fig:GFM-and-GFL-diagram}~(a). It employs either virtual synchronous generator (VSG) or droop-based power synchronization control (PSC), and uses a $1.2$~pu magnitude saturation block for current limitation (no angle prioritization). This current limiter provides accurate and fast current limitation for GFM inverters throughout power swings; however, it emulates an adaptive virtual resistance, jeopardizing the otherwise homogeneous system impedance angles in a predominantly inductive power system \cite{GFM-limiter-sat-resistive}. When the point of common coupling (PCC) voltage collapses, for example during faults or power swings, the active power reference $P^*$ is reduced. More details can be found in \cite{HJ-GFM-distance}.

Fig.~\ref{fig:GFM-and-GFL-diagram}~(b) shows the implemented GFL inverter control diagram. Synchronization to the PCC voltage is achieved using a decoupled double synchronous reference frame phase-locked loop (DDSRF-PLL). During normal operation, it utilizes the flexible positive- and negative-sequence control (FPNSC). When the PCC voltage dips during faults and power swings, it injects dynamic active and reactive currents based on the voltage level. If the PCC voltage dips below 0.5~pu, the GFL inverter is required to inject full reactive current \cite{PSB-Yongxin-GFL}. The total current is limited to $I_\mathrm{max}=1.2$~pu.

\section{Characteristics of the SCV and its Estimation}
\label{sec:SCV-characteristics}

\begin{figure}[h]
    \centering
    \includegraphics[width=0.6\linewidth]{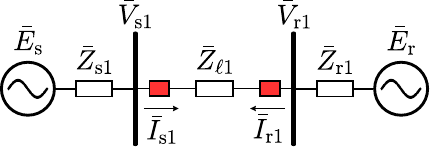}
    \caption{Two-source equivalent system with distance relays.}
    \label{fig:two-source-swing}
\end{figure}

When the faulted line in Fig.~\ref{fig:two-source-parallel-ohl} is disconnected, the equivalent system in Fig.~\ref{fig:two-source-swing} is obtained. The sending- and remote-end sources have the internal voltages $\bar{E}_\mathrm{s} = E_\mathrm{s}\angle\delta_\mathrm{s}$ and $\bar{E}_\mathrm{r} = E_\mathrm{r}\angle\delta_\mathrm{r}$, respectively. The sending-end relay (subscript ``s'') computes the positive-sequence apparent impedance 
\begin{equation}
\begin{aligned}
    \bar{Z}_1 &= \frac{\bar{V}_\mathrm{s1}}{\bar{I}_\mathrm{s1}} =\frac{\bar{E}_\mathrm{s}-\bar{I}_\mathrm{s1}\bar{Z}_\mathrm{s1}}{\bar{I}_\mathrm{s1}} = \frac{\bar{E}_\mathrm{s}}{\bar{E}_\mathrm{s}-\bar{E}_\mathrm{r}}\bar{Z}_\mathrm{sr\ell1}-\bar{Z}_\mathrm{s1},
\end{aligned} 
\label{Z1-apparent-general}
\end{equation}
where $\bar{Z}_\mathrm{sr\ell1} = \bar{Z}_\mathrm{s1}+\bar{Z}_\mathrm{r1}+\bar{Z}_\mathrm{\ell1}$ is the sum of the impedances. A time-varying angle difference $\delta(t) = \delta_\mathrm{s}(t)-\delta_\mathrm{r}(t)$ between the two equivalent sources yields a moving $\bar{Z}_1(t)$.

When $\delta = 180^\circ$, $\bar{Z}_1$ reaches the so-called electrical center (EC). Its location in the impedance plane depends on the system impedances $\bar{Z}_\mathrm{s1}$, $\bar{Z}_\mathrm{\ell1}$ and $\bar{Z}_\mathrm{r1}$, as well as the voltage magnitudes $E_\mathrm{s}$ and $E_\mathrm{r}$. If $E_\mathrm{s}=E_\mathrm{r}$, it holds that \cite{PSB-impact-of-wind}
\begin{equation}
    \bar{Z}_1(\delta(t)) = \frac{\bar{Z}_\mathrm{sr\ell1}}{2} \Bigg( 1-j\cot\bigg(\frac{\delta(t)}{2} \bigg) \Bigg)-\bar{Z}_\mathrm{s1}
    \label{Z1-delta},
\end{equation}
where the trajectory of $\bar{Z}_1$ follows along a straight line that intersects $\bar{Z}_\mathrm{sr\ell1}$ orthogonally at its middle point. When $\bar{Z}_\mathrm{s1}=\bar{Z}_\mathrm{r1}$, the EC is located at the middle of the transmission line, i.e. $\bar{Z}_1(180^\circ) = 0.5\bar{Z}_\mathrm{\ell 1}$. If $E_\mathrm{s}\neq E_\mathrm{r}$, the EC shifts toward the lower voltage source and $\bar{Z}_1$ travels along a circle \cite{PSB-Tutorial-Normann}. 

The SCV is defined as the voltage at the impedance location where the voltage becomes zero when $\delta(t) = 180^\circ$, i.e. it is the voltage at the EC. If as seen from a certain bus, a zero voltage exists close to the transmission line impedance during a power swing, the situated distance relay will perceive it as an almost bolted three-phase fault and trip unless blocked.

\subsection{Characteristics of the SCV}
\begin{figure}[h]
    \centering
    \includegraphics[width=1\linewidth]{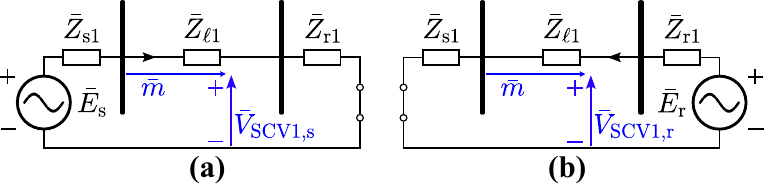}
    \caption{Source contributions to the SCV: \textbf{(a)} sending end; \textbf{(b)} remote end.}
    \label{fig:SCV-solve-super}
\end{figure}

Similar to how the dual-blinder PSB/OST schemes use the positive-sequence apparent impedance $\bar{Z}_1$, the positive-sequence SCV will be considered. Positive-sequence quantities average out the balanced oscillations occurring in each phase from the sliding-window discrete Fourier transform (DFT) during power swings \cite{PSB-Tutorial-Normann}. Let the two voltage sources in Fig.~\ref{fig:two-source-swing} be $e_\mathrm{s}(t) = E_\mathrm{s}\cos(\omega t + \delta)$ and $e_\mathrm{r}(t)=E_\mathrm{r}\cos(\omega t)$. The SCV location is assumed to be at a relative line length $\bar{m}$ from the sending-end source. Note that $\bar{m}$ is a complex number because the EC is not necessarily on the protected line. In that case, the SCV denoted $\bar{V}_\mathrm{SCV1} = \bar{V}_\mathrm{SCV1,s}+\bar{V}_\mathrm{SCV1,r}$ can be solved from Fig.~\ref{fig:SCV-solve-super} according to
\begin{equation}
    \begin{aligned}
        &\bar{V}_\mathrm{SCV1} = \bar{E}_\mathrm{s}\Bigg[\frac{(1-\bar{m})\bar{Z}_\mathrm{\ell1}+\bar{Z}_\mathrm{r1}}{\bar{Z}_\mathrm{sr\ell1}} \Bigg] + \bar{E}_\mathrm{r}\Bigg[\frac{\bar{Z}_\mathrm{s1}+\bar{m}\bar{Z}_\mathrm{\ell1}}{\bar{Z}_\mathrm{sr\ell1}}\Bigg] \\
        &= \bar{E}_\mathrm{s}\bar{m}_\mathrm{r} +\bar{E}_\mathrm{r}\bar{m}_\mathrm{s} \\
        &= E_\mathrm{s}\bar{m}_\mathrm{r}e^{j\delta(t)} + E_\mathrm{r}\bar{m}_\mathrm{s}.
    \end{aligned}
    \label{V-SCV1-phasor}
\end{equation}
From the definition that $\bar{V}_\mathrm{SCV1} =0$ and $\bar{E}_\mathrm{s} = -E_\mathrm{s}$ when $\delta=180^\circ$, it further holds that
\begin{equation}
    -E_\mathrm{s}\bar{m}_\mathrm{r} + E_\mathrm{r}\bar{m}_\mathrm{s} = 0.
    \label{SCV-formula-180deg}
\end{equation}
Solving for the SCV location $\bar{m}$ gives
\begin{equation}
    \bar{m} = \frac{E_\mathrm{s}\bar{Z}_\mathrm{\ell1}-E_\mathrm{r}\bar{Z}_\mathrm{s1}+E_\mathrm{s}\bar{Z}_\mathrm{r1}}{\bar{Z}_\mathrm{\ell1}(E_\mathrm{s}+E_\mathrm{r})},
\end{equation}
which is the EC of (\ref{Z1-apparent-general}) in terms of $\bar{Z}_1(180^\circ) = \bar{m}\bar{Z}_\mathrm{\ell 1}$. If $\bar{m}$ is approximately real and $0\leq |\bar{m}|\leq 1$, the sending-end distance zone 1 eventually sees a bolted three-phase fault at approximately distance $|\bar{m}|$ on its protected line during the power swing. If $E_\mathrm{s}=E_\mathrm{r}=E$, the SCV location simplifies to
\begin{equation}
    \bar{m} = \frac{\bar{Z}_\mathrm{\ell1}-\bar{Z}_\mathrm{s1}+\bar{Z}_\mathrm{r1}}{2\bar{Z}_\mathrm{\ell1}},
    \label{m-SCV-EsErE}
\end{equation}
and for the special case when the source impedances are equal, the SCV location becomes $\bar{m}=1/2$ (at the middle of $\bar{Z}_\mathrm{\ell 1}$).

Using that $E_\mathrm{s}\bar{m}_\mathrm{r} = E_\mathrm{r}\bar{m}_\mathrm{s}$ (\ref{SCV-formula-180deg}), the expression for $\bar{V}_\mathrm{SCV1}$ in (\ref{V-SCV1-phasor}) can be reformulated into two equivalent expressions:
\begin{equation}
    \bar{V}_\mathrm{SCV1} =
    \begin{cases}
        E_\mathrm{s}\bar{m}_\mathrm{r}\Big( e^{j\delta}+1\Big)\\ E_\mathrm{r}\bar{m}_\mathrm{s}\Big(e^{j\delta}+1 \Big),
        \end{cases}
        \label{VSCV1-piecewise1}
\end{equation}
where $\bar{m}_\mathrm{s}+\bar{m}_\mathrm{r} = 1$ by definition of the voltage division in (\ref{V-SCV1-phasor}). Assuming the two sources in Fig.~\ref{fig:two-source-swing} have an equal voltage magnitude $E_\mathrm{s}=E_\mathrm{r} = E$, the SCV location $\bar{m}$ is given by (\ref{m-SCV-EsErE}). Substituting this expression for $\bar{m}$ into (\ref{V-SCV1-phasor}), the two factors $\bar{m}_\mathrm{s}$ and $\bar{m}_\mathrm{r}$ can be solved. For instance, $\bar{m}_\mathrm{r}$ becomes 
\begin{equation}
    \bar{m}_\mathrm{r} = \frac{\bar{Z}_\mathrm{s1}+\bar{m}\bar{Z}_\mathrm{\ell1}}{\bar{Z}_\mathrm{sr\ell1}} = \frac{\bar{Z}_\mathrm{s1}+\bar{Z}_\mathrm{r1}+\bar{Z}_\mathrm{\ell1}}{2\bar{Z}_\mathrm{sr\ell1}} = \frac{1}{2},
\end{equation}
which gives $\bar{m}_\mathrm{s} = 1-\bar{m}_\mathrm{r} = 1/2$. Hence, when the two sources have an equal voltage magnitude, $\bar{V}_\mathrm{SCV1}$ in (\ref{VSCV1-piecewise1}) can be reformulated into the single expression
\begin{equation}
\begin{aligned}
\bar{V}_\mathrm{SCV1} &= 
        \frac{E}{2}\bigg(e^{j\delta}+1 \bigg)
        =E\cos\bigg(\frac{\delta}{2} \bigg)e^{j\delta/2} = \hat{V}_\mathrm{SCV1}e^{j\delta/2},
\end{aligned}
\label{VSCV1}
\end{equation}
and becomes in the time domain equal to
\begin{equation}
\begin{aligned}
    v_\mathrm{SCV1}(t) &=\hat{V}_\mathrm{SCV1}\cos\bigg(\omega t  + \frac{\delta}{2}\bigg).
\end{aligned} 
\label{vSCV1}
\end{equation}
As can be seen, $v_\mathrm{SCV1}(t)$ is an amplitude-modulated sinusoidal waveform. Under normal load conditions when the two sources in Fig.~\ref{fig:two-source-swing} are synchronized, the SCV amplitude $\hat{V}_\mathrm{SCV1}$ takes on a constant value close to $E$ because $\cos(\delta/2)\approx1$ for practical load angles.

It can be concluded from (\ref{vSCV1}) that under the condition that $E_\mathrm{s}= E_\mathrm{r}$, the SCV has upper and lower bounds, and is independent of the line and source impedances because no assumptions of them were made when solving for $v_\mathrm{SCV1}(t)$. Moreover, the instantaneous amplitude $\hat{V}_\mathrm{SCV1}$ is directly related to $\delta$ and $\hat{V}_\mathrm{SCV1} = 0$ when $\delta = 180^\circ$. Time-varying source impedances change the EC but not the SCV characteristics. The rate of change of $\hat{V}_\mathrm{SCV1}(\delta(t))$ becomes
\begin{equation}
    \frac{d\hat{V}_\mathrm{SCV1}}{dt} = -\frac{E}{2}\sin\bigg(\frac{\delta(t)}{2} \bigg)\frac{d\delta}{dt},
    \label{SCV-derivative}
\end{equation}
which only depends on $\delta(t)$ and $d\delta/dt$. With a constant slip angular velocity $\omega_\mathrm{slip}=2\pi f_\mathrm{slip}$, the rate of change of the SCV amplitude becomes a sinusoid, with $\delta(t) = \omega_\mathrm{slip}t$. 

\subsection{Locally Estimating the SCV}

$\hat{V}_\mathrm{SCV1}$ cannot be directly computed by the distance relays in Fig.~\ref{fig:two-source-swing} because $\delta$ is unknown. For the sending-end relay, an estimation can be made by projecting the positive-sequence voltage $\bar{V}_\mathrm{s1}$ onto the positive-sequence current $\bar{I}_\mathrm{s1}$ \cite{PSB-Tutorial-Normann}:
\begin{equation}
\begin{aligned}
    &\bar{V}_\mathrm{SCV1} \approx \frac{P_\mathrm{s}}{|\bar{I}_\mathrm{s1}|}e^{j\arg\{\bar{I}_\mathrm{s1}\}} =\underbrace{| \bar{V}_\mathrm{s1}|\cos(\varphi_\mathrm{s})}_{V_\mathrm{SCV1}}e^{j\arg\{\bar{I}_\mathrm{s1}\}},
\end{aligned} 
\label{SCV-approx-eq}
\end{equation}
where $P_\mathrm{s}$ is the sending-end active power, $\varphi_\mathrm{s} = \arg\{\bar{V}_\mathrm{s1}\}-\arg\{\bar{I}_\mathrm{s1}\}$ is the power factor angle and $V_\mathrm{SCV1} =|\bar{V}_\mathrm{s1}|\cos(\varphi_\mathrm{s})$ is used to approximate $\hat{V}_\mathrm{SCV1}$. The time derivative of $V_\mathrm{SCV1}$ is then numerically computed to estimate $d\hat{V}_\mathrm{SCV1}/dt$. 
 
For the true instantaneous amplitude $\hat{V}_\mathrm{SCV1}$ (\ref{VSCV1}) and rate of change $d\hat{V}_\mathrm{SCV1}/dt$ (\ref{SCV-derivative}), the defined range of $\delta$ does not matter. However, when relating $\hat{V}_\mathrm{SCV1}$ directly to $V_\mathrm{SCV1}$ from (\ref{SCV-approx-eq}), the defined range of $\delta$ matters. This is because when $P_\mathrm{s}$ reverses in Fig.~\ref{fig:two-source-swing}, $\cos(\varphi_\mathrm{s})$ changes sign. The relation between $\delta$ and $\varphi_\mathrm{s}$ can be mathematically formulated noting that $\angle\bar{Z}_1 = \angle\bar{V}_\mathrm{s1}-\angle\bar{I}_\mathrm{s1} = \varphi_\mathrm{s}$ and $\tan(\varphi_\mathrm{s}) = \text{Im}(\bar{Z}_1)/\text{Re}(\bar{Z}_1)$. Hence, splitting up (\ref{Z1-delta}) into its real and imaginary parts gives 
\begin{equation}
    \tan(\varphi_\mathrm{s}) = \frac{\text{Im}(\bar{Z}_1)}{\text{Re}(\bar{Z}_1)} = \frac{-R_\mathrm{sr\ell 1}\cot(\delta/2) + X_\mathrm{sr\ell 1} - 2X_\mathrm{s1}}{X_\mathrm{sr\ell 1}\cot(\delta/2) + R_\mathrm{sr\ell 1} - 2R_\mathrm{s1}}.
    \label{tan(phi-s)-general}
\end{equation}
In a highly inductive system with minor losses ($R_\mathrm{sr\ell 1} \approx 0$) and strong sources with high SIRs, the expression simplifies to $\tan(\varphi_\mathrm{s}) = Q_\mathrm{s}/P_\mathrm{s} =\tan(\delta/2)$, giving the explicit relationship $\varphi_\mathrm{s} = \delta/2$ (also $\varphi_\mathrm{r} = -\delta/2$). Hence, defining $\delta\in[0, 2\pi)$ for $\hat{V}_\mathrm{SCV1}$ in (\ref{VSCV1}) gives a correct comparison with $V_\mathrm{SCV1}$ in (\ref{SCV-approx-eq}). For example, if $\delta$ starts from 0 and rotates counterclockwise ($d\delta/dt>0$), the locally estimated signed projection $V_\mathrm{SCV1}$ decreases from 1~pu to 0 when $\delta$ goes from 0 to $\pi$. As $\delta$ continues, $P_\mathrm{s}$ reverses and $\text{Re}(\bar{Z}_1)$ crosses 0, making $V_\mathrm{SCV1}$ continue downwards from 0 to -1~pu. Hence, $V_\mathrm{SCV1}$ behaves as $\hat{V}_\mathrm{SCV1}$ does for $\delta \in [0, 2\pi)$. Moreover, similar to $d\hat{V}_\mathrm{SCV1}/dt$ (\ref{SCV-derivative}), $dV_\mathrm{SCV1}/dt$ directly mirrors the rotational direction of $\delta$ when $\delta \in [0, 2\pi)$. If instead $\delta\in[-\pi, \pi)$, $\varphi\in[-\pi/2, \pi/2)$ makes $\cos(\varphi_\mathrm{s})$ positive over a complete cycle of $\delta$, which does not account for the reversal of $P_\mathrm{s}$ when relays compute $V_\mathrm{SCV1}$.

\begin{figure}[h]
    \centering
    \includegraphics[width=0.9\linewidth]{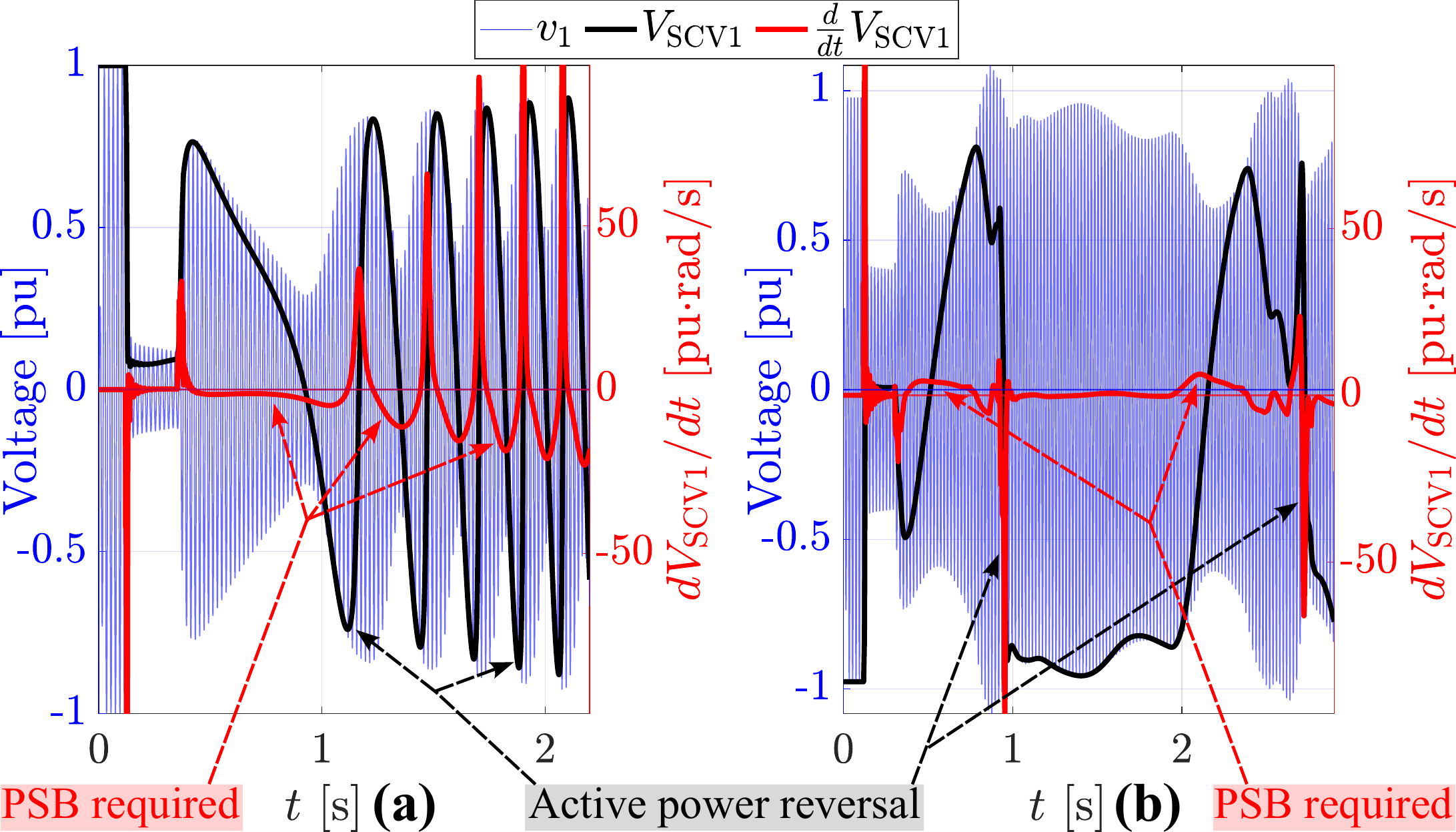}
    \caption{SCV locally estimated: \textbf{(a)} $\delta(t)$ rotating counterclockwise in an SG-based system; \textbf{(b)} $\delta(t)$ rotating clockwise in a multi-source system.}
    \label{fig:SCV-local}
\end{figure}

It should be pointed out that, although $\delta$ rotating counterclockwise or clockwise implies that $dV_\mathrm{SCV1}/dt$ only takes on negative or positive values for $\delta\in[0, 2\pi)$, respectively, it experiences discontinuities when $P_\mathrm{s}$ reverses and $\bar{Z}_1$ crosses the reactive axis. Unless $\delta(t)$ causes $\bar{Z}_1$ to oscillate back and forth around $\text{Re}(\bar{Z}_1)=0$ without completing a swing cycle, this occurs only once for a complete swing cycle. Fig.~\ref{fig:SCV-local} illustrates the impact of $P_\mathrm{s}$ reversal on $V_\mathrm{SCV1}$ for two out-of-step (OOS) conditions from PSCAD, causing large $dV_\mathrm{SCV1}/dt$ values and of opposite sign compared to the rotation of $\delta$. In Fig.~\ref{fig:SCV-local}~(a), the positive spikes (other than the fault clearance) are due to $P_\mathrm{s}$ reversal and occur when $\text{Re}(\bar{Z}_1)$ transitions from negative to positive (going into the forward load region). In Fig.~\ref{fig:SCV-local}~(b), the negative spikes (other than the fault clearance) are due to $\text{Re}(\bar{Z}_1)$ transitioning from positive to negative.

\section{Impacts of IBRs on the SCV Estimation}
\label{sec:Impact-IBRs-SCV}
The SCV approximation $V_\mathrm{SCV1}$ from (\ref{SCV-approx-eq}) holds well with SGs, but unlike $\hat{V}_\mathrm{SCV1}$ from (\ref{VSCV1}), it depends on the system impedances and power flow on the line; it is thus impacted by the control of IBRs. GFM inverters emulate dynamic source impedances during power swings to limit the current injection and GFL inverters behave as current sources controlled to inject a specific amount of active and reactive currents within the overcurrent headroom. In this section, these impacts on the local SCV estimation are analyzed. As motivated, $\delta\in [0, 2\pi)$ will be used, and without loss of generality, counterclockwise power swings ($d\delta/dt>0$) will be considered in Fig.~\ref{fig:two-source-swing}. 

\subsection{Impacts of GFM Inverters on the SCV Estimation}
\begin{figure}[h]
    \centering
    \includegraphics[width=1\linewidth]{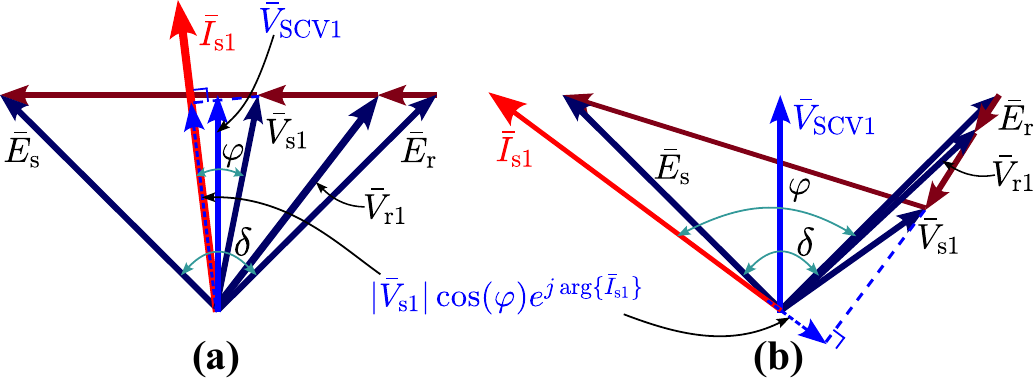}
    \caption{True and estimated SCV phasors when $\delta=90^\circ$: \textbf{(a)} GFM inverter with high inductive $\bar{Z}_\mathrm{s1}$; \textbf{(b)} GFM inverter with high resistive $\bar{Z}_\mathrm{s1}$.}
    \label{fig:SCV-approx-drawing}
\end{figure}

The limited overcurrent capability of GFM inverters reduces $|\bar{V}_\mathrm{s1}|$ and impacts $\varphi_\mathrm{s}$. Fig.~\ref{fig:SCV-approx-drawing} illustrates the accuracy of (\ref{SCV-approx-eq}) in the two-source equivalent system in Fig.~\ref{fig:two-source-swing} when $\bar{Z}_\mathrm{s1}$ is significantly larger in magnitude than $\bar{Z}_\mathrm{\ell 1}$ and $\bar{Z}_\mathrm{r1}$, and either inductive or resistive. The brown vectors are the voltage drops over the three impedances from $\bar{E}_\mathrm{r}$ to $\bar{E}_\mathrm{s}$. The SIR of the remote source is 0.3 (typical for long lines \cite{SIR-tutorial}), $\delta=90^\circ$ and $\bar{Z}_\mathrm{\ell 1}$ is obtained from the benchmark PSCAD system explained in Section~\ref{sec:benchmark-performance-eval}. MATLAB is used to compute the phasor diagrams from Fig.~\ref{fig:two-source-swing}. As can be seen in Fig.~\ref{fig:SCV-approx-drawing}~(a), the SCV estimation (\ref{SCV-approx-eq}) holds well if $\bar{Z}_\mathrm{s1}$ is of high magnitude but remains inductive. However, when $\bar{Z}_\mathrm{s1}$ is predominantly resistive, the accuracy starts to wane, as shown in Fig.~\ref{fig:SCV-approx-drawing}~(b). 

\begin{figure}[h]
    \centering
    \includegraphics[width=0.75\linewidth]{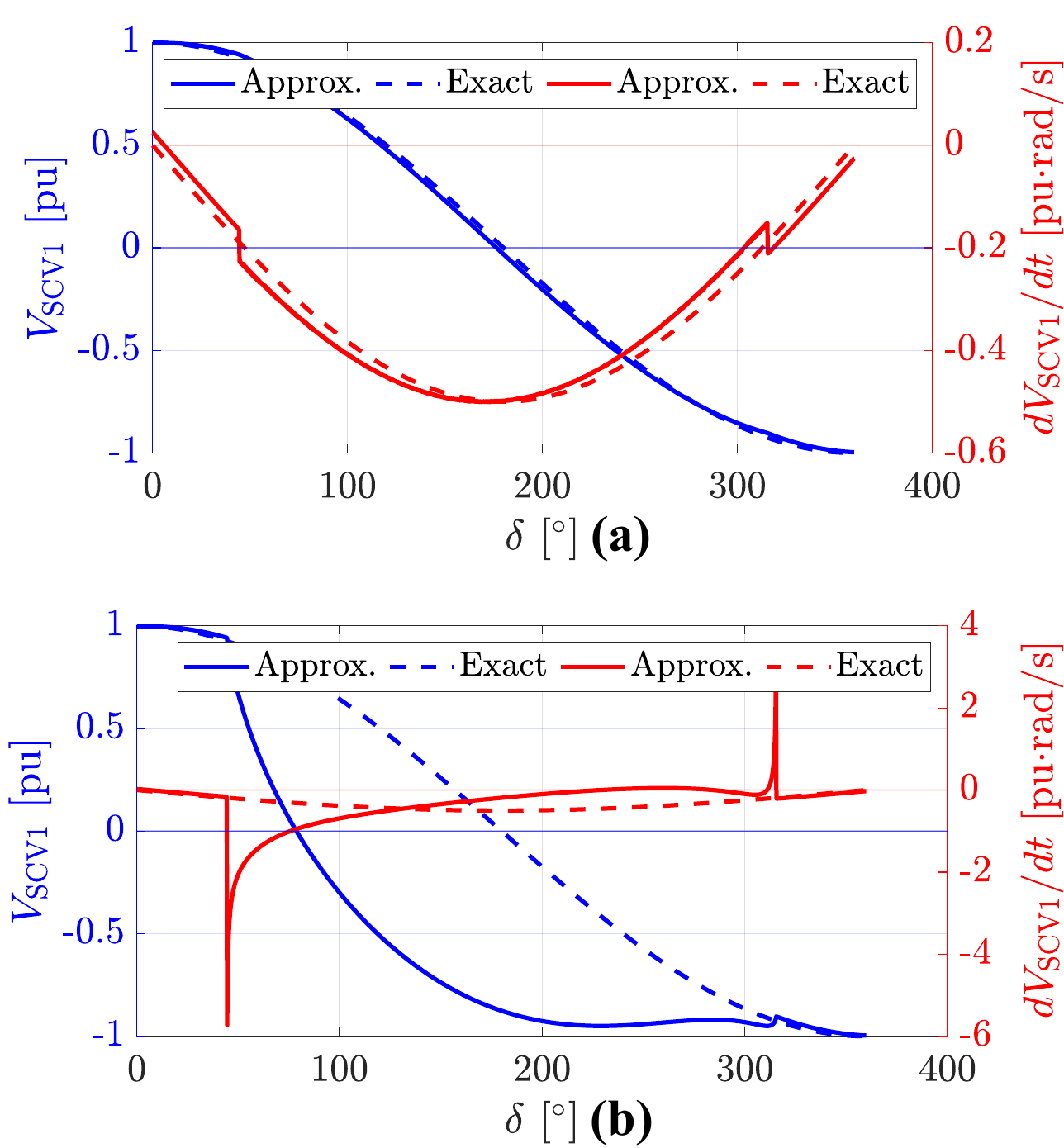}
    \caption{Characteristics of the SCV and its rate of change: \textbf{(a)} GFM inverter with inductive current limiter; \textbf{(b)} GFM inverter with resistive current limiter.}
    \label{fig:SCV-approx}
\end{figure}

To get an understanding of how a changing source impedance for current limitation by GFM inverters impacts $V_\mathrm{SCV1}$ and its time derivative, parameter sweeps are conducted over a complete cycle of $\delta$ on the system in Fig.~\ref{fig:two-source-swing}, which starts with $\bar{Z}_\mathrm{s1}/\bar{Z}_\mathrm{\ell 1}=\bar{Z}_\mathrm{r1}/\bar{Z}_\mathrm{\ell 1} = 0.3$. However, for all values of $\delta$ when $|\bar{I}_\mathrm{s1}|>1.2$~pu, which happens around $\delta=45^\circ$, an additional source impedance $\bar{Z}_\mathrm{VI1}$ (inductive or resistive) is added at the sending-end to limit the injected current to $1.2$~pu. Fig.~\ref{fig:SCV-approx} shows how the estimated $V_\mathrm{SCV1}$ (\ref{SCV-approx-eq}) and numerically computed $dV_\mathrm{SCV1}/dt$ compares with the true quantities $\hat{V}_\mathrm{SCV1}$ (\ref{VSCV1}) and $d\hat{V}_\mathrm{SCV1}/dt$ (\ref{SCV-derivative}). Here, $d\delta/dt = \omega_\mathrm{slip}=1$~rad/s, meaning that $f_\mathrm{slip} \approx 0.16$~Hz. All voltage quantities have been normalized to per-unit values. As can be seen, the approximated $V_\mathrm{SCV1}$ and $dV_\mathrm{SCV1}/dt$ are accurate when the source impedance is inductive, but differences occur when it is resistive. The characteristics of $V_\mathrm{SCV1}$ and $dV_\mathrm{SCV1}/dt$ seen in Fig.~\ref{fig:SCV-approx} will be explained mathematically in the next sections.

\subsubsection{Inductive GFM Inverter Current Limiters}
GFM inverters may use an inductive virtual impedance (VI) in the outer voltage control loop to limit the current by emulating an additional inductive contribution to $\bar{Z}_\mathrm{s1}$, as in Fig.~\ref{fig:SCV-approx-drawing}~(a). Hence, $X_\mathrm{s1}(\delta(t))$ varies with time. Assuming the system in Fig.~\ref{fig:two-source-swing} is predominantly inductive ($R_\mathrm{sr\ell 1}\approx 0$), (\ref{tan(phi-s)-general}) simplifies to
\begin{equation}
    \tan(\varphi_\mathrm{s}) = \bigg(1-\frac{2X_\mathrm{s1}(\delta(t))}{X_\mathrm{s1}(\delta(t))+X_\mathrm{\ell 1}+X_\mathrm{r1}}\bigg)\tan(\delta(t)/2).
    \label{tan(phi-s)-inductive}
\end{equation}
It can be noted that a sign flip of $\varphi_\mathrm{s}$ occurs when $2X_\mathrm{s1}>X_\mathrm{sr\ell 1}$, but this has no impact on the power factor $\cos(\varphi_\mathrm{s})$ which is an even function. The sign flip occurs because the high $X_\mathrm{s1}$ moves $\bar{I}_\mathrm{s1}$ from lagging to leading $\bar{V}_\mathrm{s1}$, such as in Fig.~\ref{fig:SCV-approx-drawing}~(a). However, as long as the system impedances remain inductive (homogeneous angles), any $X_\mathrm{s1}(\delta(t))$ change in $\bar{Z}_\mathrm{s1}$ shifts $\bar{V}_\mathrm{s1}$ orthogonally to $\bar{I}_\mathrm{s1}$, making the orthogonal projection of $\bar{V}_\mathrm{s1}$ onto $\bar{I}_\mathrm{s1}$ still accurate in (\ref{SCV-approx-eq}). Mathematically, $\bar{V}_\mathrm{s1} = \bar{E}_\mathrm{s}-\bar{Z}_\mathrm{s1}\bar{I}_\mathrm{s1}$ grows by the factor $1/\cos(\varphi_\mathrm{s})$, yielding
\begin{equation}
    V_\mathrm{SCV1} = |\bar{V}_\mathrm{s1}|\cos(\varphi_\mathrm{s}) = \bigg|\frac{E\cos(\delta/2)}{\cos(\varphi_\mathrm{s})}\bigg|\cos(\varphi_\mathrm{s}) = \hat{V}_\mathrm{SCV1}. 
\end{equation}
The last equality holds when $\delta\in[0, 2\pi)$. Consequently, the local SCV estimation (\ref{SCV-approx-eq}) holds well with GFM inverters emulating an inductive source impedance change to limit the current. In Fig.~\ref{fig:SCV-approx}~(a) the current limiter has a slight impact because the system impedances have a small amount of resistance while the VI is purely inductive.

\subsubsection{Resistive GFM Inverter Current Limiters}
The current limiter of the implemented GFM inverter in Fig.~\ref{fig:GFM-and-GFL-diagram}~(a) emulates an additional source resistance to curtail the current \cite{GFM-limiter-sat-resistive}. Assuming $\bar{Z}_\mathrm{r1} \approx jX_\mathrm{r1}$ and $\bar{Z}_\mathrm{\ell 1} \approx jX_\mathrm{\ell 1}$, (\ref{tan(phi-s)-general}) simplifies to
\begin{equation}
    \tan(\varphi_\mathrm{s}) = \frac{-R_\mathrm{s1}(\delta(t))\cot(\delta(t)/2) + X_\mathrm{\ell 1}+ X_\mathrm{r1} - X_\mathrm{s1}}{X_\mathrm{sr\ell 1}\cot(\delta(t)/2) - R_\mathrm{s1}(\delta(t))}.
    \label{tan(phi-s)-resistive}
\end{equation}
It can be seen in Fig.~\ref{fig:SCV-approx}~(b) that $dV_\mathrm{SCV1}/dt$ accelerates when the current limiter gets activated around $\delta\approx 45^\circ$ and it stalls from approximately $\delta\approx 180^\circ$ (the EC) to $ \delta\approx  315^\circ$. To understand why this happens, consider that $\bar{V}_\mathrm{s1} = \bar{E}_\mathrm{r}+j(X_\mathrm{r1} + X_\mathrm{\ell 1}) \bar{I}_\mathrm{s1}$. Because the inductive voltage drop $j(X_\mathrm{r1} + X_\mathrm{\ell 1})\bar{I}_\mathrm{s1}$ is perpendicular to $\bar{I}_\mathrm{s1}$, it contributes zero projection onto $\bar{I}_\mathrm{s1}$. Therefore, the local SCV estimation simplifies directly to the projection of $\bar{E}_\mathrm{r} = E$ onto $\bar{I}_\mathrm{s1}$:
\begin{equation}
    V_\mathrm{SCV1}=|\bar{V}_\mathrm{s1}|\cos(\varphi_\mathrm{s}) \approx E\cos(\angle\bar{I}_\mathrm{s1}),
    \label{VSC1-GFM-sat}
\end{equation}
and with $\bar{E}_\mathrm{s}=Ee^{j\delta}$, the sending-end current is
\begin{equation}
\begin{aligned}
    \bar{I}_\mathrm{s1} &= \frac{\bar{E}_\mathrm{s}-\bar{E}_\mathrm{r}}{R_\mathrm{s1}(\delta(t))+jX_\mathrm{sr\ell 1}} \\
    &= \frac{2E\sin(\delta(t)/2)}{Z_\mathrm{sr\ell 1}(\delta(t))}\exp\bigg(j\frac{\delta(t)}{2} +\frac{\pi}{2} - \theta_\mathrm{sr\ell 1}(\delta(t))\bigg),
\end{aligned} 
\label{Is1-GFM-sat}
\end{equation}
where $Z_\mathrm{sr\ell 1}(\delta(t)) = |R_\mathrm{s1}(\delta(t))+jX_\mathrm{sr\ell 1}|$ and $\theta_\mathrm{sr\ell 1}(\delta(t)) = \arg[R_\mathrm{s1}(\delta(t))+jX_\mathrm{sr\ell 1}] = \arctan[X_\mathrm{sr\ell 1}/R_\mathrm{s1}(\delta(t))]$. Before current limitation is needed, $\theta_\mathrm{sr\ell 1}(\delta(t))\approx 90^\circ$ and $\angle\bar{I}_\mathrm{s1}\approx \delta(t)/2$, making $V_\mathrm{SCV1}=E\cos(\delta(t)/2) = \hat{V}_\mathrm{SCV1}$. When the current limiter engages for approximately $\delta \in[45^\circ, 315^\circ]$, $R_\mathrm{s1}(\delta(t))$ dynamically adjusts to clamp $|\bar{I}_\mathrm{s1}|=1.2$~pu, forcing $\theta_\mathrm{sr\ell 1}(\delta(t))$ to vary with $\delta$ in opposite directions before and after $\bar{Z}_1$ reaches the EC at $\delta=180^\circ$:
\begin{itemize}
    \item When $\delta\in [45^\circ, 180^\circ)$, the driving voltage $|\bar{E}_\mathrm{s}-\bar{E}_\mathrm{r}|=2E\sin(\delta(t)/2)$ increases. To keep $|\bar{I}_\mathrm{s1}|=1.2$~pu, $R_\mathrm{s1}$ increases and $\theta_\mathrm{sr\ell 1}$ decreases from $90^\circ$ towards $0^\circ$. Consequently, $\angle\bar{I}_\mathrm{s1}$ sweeps rapidly past $90^\circ$, causing the steep negative slope $dV_\mathrm{SCV1}/dt$ and premature zero crossing of $V_\mathrm{SCV1}$ in Fig.~\ref{fig:SCV-approx}~(b). 
    \item When $\delta\in [180^\circ, 315^\circ)$, the driving voltage decreases with $\delta$, so the current limiter gradually reduces $R_\mathrm{s1}$ to maintain $|\bar{I}_\mathrm{s1}|=1.2$~pu. During this process, $\theta_\mathrm{sr\ell 1}$ increases back towards its $90^\circ$ inductive baseline. The growth of $\theta_\mathrm{sr\ell 1}$ subtracts from the advance of $\delta(t)/2$, stalling $\angle\bar{I}_\mathrm{s1}$, resulting in the $V_\mathrm{SCV1}$ plateau followed by the positive $dV_\mathrm{SCV1}/dt$ spike seen in Fig.~\ref{fig:SCV-approx}~(b).  
\end{itemize}

The premature zero crossing of $V_\mathrm{SCV1}$ in Fig.~\ref{fig:SCV-approx} can be calculated by solving for $\delta_0$ from $\text{Re}[\bar{Z}_1(\delta_0)]=0$ (\ref{Z1-delta}). Using the denominator in (\ref{tan(phi-s)-resistive}) gives
\begin{equation}
    \delta_0 = 2\text{arccot}\bigg(\frac{R_\mathrm{s1}}{X_\mathrm{sr\ell 1}}\bigg) = \pi - 2\arctan\bigg(\frac{R_\mathrm{s1}}{X_\mathrm{sr\ell 1}}\bigg).
\end{equation}
In conventional power systems, $R_\mathrm{s1}/X_\mathrm{sr\ell 1}\approx 0$, giving $\delta_0 \approx \pi$ (same as the true SCV). However, the GFM inverter in Fig.~\ref{fig:GFM-and-GFL-diagram}~(a) causes $R_\mathrm{s1}$ to increase and even become substantially larger than $X_\mathrm{sr\ell 1}$, moving $\delta_0$ to a lower value before $\bar{Z}_1$ reaches the EC. Moreover, the EC (\ref{m-SCV-EsErE}) becomes equal to $\bar{Z}_\mathrm{1}(180^\circ)=\bar{m}\bar{Z}_\mathrm{\ell 1} = 0.5[j(X_\mathrm{\ell 1}+X_\mathrm{r1}-X_\mathrm{s1})-R_\mathrm{s1}(\delta(t))]$. Consequently, the EC location is dynamically pushed leftwards in the impedance plane into the second quadrant, causing $\bar{Z}_1$ to travel below and then behind the distance zones as $\delta$ increases towards $180^\circ$. Since $|dV_\mathrm{SCV1}/dt|$ accelerates by $R_\mathrm{s1}(\delta(t))$ before $\delta=180^\circ$, $|d\bar{Z}_1/dt|$ also accelerates since it is related to $\varphi_\mathrm{s}$. This occurs for the range of $\delta$ when $\bar{Z}_1$ typically crosses the blinders in a swing cycle, i.e. when blinder-based PSB schemes need to pickup. Hence, $R_\mathrm{s1}(\delta(t))$ causes severe interoperability issues with blinder-based PSB schemes that rely on a slow $|d\bar{Z}_1/dt|$ precisely when $\bar{Z}_1$ crosses them. Moreover, because the EC is pushed leftwards in the impedance plane, there is an increased risk that $\bar{Z}_1$ immediately enters distance zone 1 after a zone 2 fault is cleared on the healthy line in Fig.~\ref{fig:two-source-parallel-ohl}. Blinder-based PSB schemes cannot prevent tripping during such $\bar{Z}_1$ fault-to-swing transitions. This can occur over a wide range of initial values of $\delta$ with GFM inverters having a saturation magnitude current limiter---both stable and unstable power swings \cite{HJ-PSB-IQ2}.

\subsection{Impacts of GFL Inverters on the SCV Estimation}

GFL inverters track $\bar{V}_\mathrm{s1}$ and directly regulate the injected active and reactive currents depending on the voltage level; they thus act as voltage-dependent current sources. Because $V_\mathrm{SCV1}$ (\ref{SCV-approx-eq}) depends on the terminal power factor angle $\varphi_\mathrm{s}$, the GFL inverter current controller in Fig.~\ref{fig:GFM-and-GFL-diagram}~(b) directly shapes $V_\mathrm{SCV1}$. Hence, the GFL inverter is also expected to accelerate $|dV_\mathrm{SCV1}/dt|$ compared to the true SCV rate of change $d\hat{V}_\mathrm{SCV1}/dt$ (\ref{SCV-derivative}). It should be emphasized that $V_\mathrm{SCV1} = 0$ only when $\varphi_\mathrm{s}=\pm 90^\circ$, which may happen shortly after the GFL inverter enters its full reactive current mode ($|\bar{V}_\mathrm{s1}|<0.5$~pu). If the PLL is locked onto $\bar{V}_\mathrm{s1}=|\bar{V}_\mathrm{s1}|e^{j\theta_\mathrm{s1}}$, the voltage phasor lies along the $d$-axis so that $\bar{V}_\mathrm{s1}=v_{\mathrm{s}d}+jv_{\mathrm{s}q} = |\bar{V}_\mathrm{s1}| + j0$ in the synchronous $dq$ reference frame ($\theta_\mathrm{s1} = \theta_\mathrm{PLL}$). With $\bar{I}_\mathrm{s1} = i_{\mathrm{s}d} + ji_{\mathrm{s}q}$, the complex power becomes
\begin{equation}
    \bar{S}_\mathrm{s} = \bar{V}_\mathrm{s1}\bar{I}_\mathrm{s1}^* = P_\mathrm{s} + jQ_\mathrm{s} = |\bar{V}_\mathrm{s1}|(i_{\mathrm{s}d}-ji_{\mathrm{s}q}),
\end{equation}
and injecting full reactive power ($Q_\mathrm{s}>0$) requires the GFL inverter setting $i_{\mathrm{s}d}=0$ and $i_{\mathrm{s}q}=-I_\mathrm{max}$. In the stationary reference frame, $\bar{V}_\mathrm{s1}=|\bar{V}_\mathrm{s1}|e^{j\theta_\mathrm{s1}}$ leads $\bar{I}_\mathrm{s1} = I_\mathrm{max}e^{j(\theta_\mathrm{s1}-90^\circ)} = -jI_\mathrm{max}e^{j\theta_\mathrm{s1}}$ by $90^\circ$, resulting in $V_\mathrm{SCV1}=0$. 

During unstable power swings, the PLL loses synchronism and tries to follow $\bar{V}_\mathrm{s1}=|\bar{V}_\mathrm{s1}|e^{j\theta_{s1}}$. A dynamic PLL phase slip may be defined from 0 to $2\pi$ as $\tilde{\theta}_\mathrm{PLL} = \theta_\mathrm{PLL}-\theta_\mathrm{s1}$. Inverse transformation from the $dq$ to stationary reference frame gives $\bar{I}_\mathrm{s1} = (i_{\mathrm{s}d}+ji_{\mathrm{s}q})e^{j\theta_\mathrm{PLL}}$, and when power swings cause $|\bar{V}_\mathrm{s1}|<0.5$~pu, reactive-current priority enforces $i_{\mathrm{s}d}=0$ and $i_{\mathrm{s}q}=-I_\mathrm{max}$ in the PLL $dq$ frame, resulting in
\begin{equation}
    \angle\bar{I}_\mathrm{s1} = \angle(-jI_\mathrm{max}e^{j\theta_\mathrm{PLL}}) =\theta_\mathrm{PLL}-\pi/2.
\end{equation}
Substituting this into $\varphi_\mathrm{s}=\angle\bar{V}_\mathrm{s1}-\angle\bar{I}_\mathrm{s1}$ gives
\begin{equation}
    \varphi_\mathrm{s} = \theta_\mathrm{s1} - (\theta_\mathrm{PLL}-\pi/2) = \pi/2 - \tilde{\theta}_\mathrm{PLL},
\end{equation}
and $V_\mathrm{SCV1}$ (\ref{SCV-approx-eq}) becomes
\begin{equation}
    V_\mathrm{SCV1} = |\bar{V}_\mathrm{s1}|\cos\bigg(\frac{\pi}{2} - \tilde{\theta}_\mathrm{PLL}\bigg) = |\bar{V}_\mathrm{s1}|\sin(\tilde{\theta}_\mathrm{PLL}). 
    \label{VSCV1-GFL-theta-tilde}
\end{equation}
With $\bar{E}_\mathrm{r} = E_\mathrm{r}e^{j\delta_\mathrm{r}}$, $\bar{V}_\mathrm{s1}$ in the stationary reference frame is
\begin{equation}
    \bar{V}_\mathrm{s1} = |\bar{V}_\mathrm{s1}|e^{j\theta_\mathrm{s1}} =  E_\mathrm{r}e^{j\delta_\mathrm{r}}+X_\mathrm{r\ell 1}I_\mathrm{max}e^{j\theta_\mathrm{PLL}},
    \label{Vs1-GFL-Imax-Qprio}
\end{equation}
where $X_\mathrm{r\ell 1} = X_\mathrm{r1} + X_\mathrm{\ell 1} \approx |\bar{Z}_\mathrm{r1}+\bar{Z}_\mathrm{\ell 1}|$ (negligible resistive losses in Fig.~\ref{fig:two-source-swing}). Multiplying (\ref{Vs1-GFL-Imax-Qprio}) by $e^{-j\theta_\mathrm{PLL}}$ projects all phasors onto the PLL synchronous reference frame:
\begin{equation}
    |\bar{V}_\mathrm{s1}|e^{-j\tilde{\theta}_\mathrm{PLL}} = E_\mathrm{r}e^{-j\delta_\mathrm{slip}} + X_\mathrm{r\ell 1}I_\mathrm{max},
    \label{Vs1-GFL-Imax-Qprio-dq}
\end{equation}
where $\delta_\mathrm{slip} = \theta_\mathrm{PLL}-\delta_\mathrm{r}$ is the total slip angle between the unlocked PLL and the remote grid. Equating the imaginary $q$-axis parts directly gives that (\ref{VSCV1-GFL-theta-tilde}) simplifies to
\begin{equation}
    V_\mathrm{SCV1} = |\bar{V}_\mathrm{s1}|\sin(\tilde{\theta}_\mathrm{PLL}) =E_\mathrm{r}\sin(\delta_\mathrm{slip}).
    \label{VSCV1-GFL}
\end{equation}
This proves that even though $|\bar{V}_\mathrm{s1}|$ and $\sin(\tilde{\theta}_\mathrm{PLL})$ both vary nonlinearly during a complete slip cycle, their product equals the $q$-axis projection of the remote source voltage when the GFL inverter is in its saturated reactive-current priority mode while riding through the voltage dip. Since $E_\mathrm{r}$ is a positive value, $V_\mathrm{SCV1}$ crosses zero once at $\delta_\mathrm{slip}=180^\circ$ for a complete slip cycle (assuming $E_\mathrm{r}\neq0$), at which point $\tilde{\theta}_\mathrm{PLL}=180^\circ$ also. Additionally, equating the $d$-axis real parts of (\ref{Vs1-GFL-Imax-Qprio-dq}) gives 
\begin{equation}
|\bar{V}_\mathrm{s1}|\cos(\tilde{\theta}_\mathrm{PLL}) = E_\mathrm{r}\cos(\delta_\mathrm{slip}) + X_\mathrm{r\ell 1}I_\mathrm{max}.
\label{theta-tilde-PLL-delta-slip-realpart}
\end{equation}
Dividing the imaginary $q$-axis expressions on each side of (\ref{Vs1-GFL-Imax-Qprio-dq}) by their respective real $d$-axis expressions gives
\begin{equation}
    \tan(\tilde{\theta}_\mathrm{PLL}) = \frac{E_\mathrm{r}\sin(\delta_\mathrm{slip})}{E_\mathrm{r}\cos(\delta_\mathrm{slip})+X_\mathrm{r\ell 1}I_\mathrm{max}} = \frac{\sin(\delta_\mathrm{slip})}{\cos(\delta_\mathrm{slip})+k},
    \label{theta-tilde-PLL-delta-slip}
\end{equation}
where $k=X_\mathrm{r\ell 1}I_\mathrm{max}/E_\mathrm{r}$. For a strong grid ($0<k<1$), the numerator $\sin(\delta_\mathrm{slip})>0$ for $\delta_\mathrm{slip}\in(0^\circ, 180^\circ)$, implying that $\tilde{\theta}_\mathrm{PLL}\in(0^\circ, 180^\circ)$, because $\delta_\mathrm{slip}$ and $\tilde{\theta}_\mathrm{PLL}$ have the same sign according to (\ref{VSCV1-GFL}). The numerator reaches zero only if $\delta_\mathrm{slip}=180^\circ$, at which point the right-hand side denominator is $k-1<0$, forcing $\tilde{\theta}_\mathrm{PLL}=180^\circ$ at this exact instance. Hence, although $\tilde{\theta}_\mathrm{PLL}(t)\neq \delta_\mathrm{slip}(t)$, they pass through $180^\circ$ at the same time, i.e. when $V_\mathrm{SCV1}=0$ (\ref{VSCV1-GFL}). This means that $\theta_\mathrm{s1}=\delta_\mathrm{r}$ when $\delta_\mathrm{slip}=180^\circ$, which is expected when transferring zero $P_\mathrm{s}$ over $jX_\mathrm{r\ell1}$ to $\bar{E}_\mathrm{r}$. Moreover, since $k>0$ adds a positive offset to the right-hand side denominator, $\tilde{\theta}_\mathrm{PLL}$ is lagging $\delta_\mathrm{slip}$ before $\bar{Z}_1$ of the distance relay reaches the EC. This causes $\tilde{\theta}_\mathrm{PLL}$ to accelerate rapidly as $\bar{Z}_1$ approaches the EC, resulting in an increased $|d\bar{Z}_1/dt|$ during this part of the power swing cycle because 
\begin{equation}
    \bar{Z}_1 = \frac{|\bar{V}_\mathrm{s1}|e^{j\theta_\mathrm{s1}}}{-jI_\mathrm{max}e^{j\theta_\mathrm{PLL}}} = j\frac{|\bar{V}_\mathrm{s1}|}{I_\mathrm{max}}e^{-j\tilde{\theta}_\mathrm{PLL}}.
\end{equation}
This $\bar{Z}_1$ acceleration could cause interoperability issues with blinder-based PSB schemes. For very weak grids or large reactive current injection ($k>1$), the right-hand side denominator of (\ref{theta-tilde-PLL-delta-slip}) is positive throughout the entire $360^\circ$ swing cycle of $\delta_\mathrm{slip}$, with a minimum value of $k-1>0$ at $\delta_\mathrm{slip}=180^\circ$. Since $|\bar{V}_\mathrm{s1}|>0$, the sign of $\cos(\tilde{\theta}_\mathrm{PLL})$ is determined entirely by the sign of $E_\mathrm{r}\cos(\delta_\mathrm{slip}) + X_\mathrm{r\ell 1}I_\mathrm{max}>0$ (\ref{theta-tilde-PLL-delta-slip-realpart}). Consequently, $\tilde{\theta}_\mathrm{PLL}$ oscillates back and forth between $-90^\circ$ and $90^\circ$, and it wraps around when $\delta_\mathrm{slip}$ crosses $180^\circ$, yet $V_\mathrm{SCV1}$ still crosses zero smoothly at $\delta_\mathrm{slip} = 180^\circ$ (\ref{VSCV1-GFL}).

\section{Proposed SCV-Based PSB/OST Scheme}
\label{sec:Proposed-SCV-scheme}
The proposed SCV-based PSB scheme computes $V_\mathrm{SCV1}$ (\ref{SCV-approx-eq}) using a sliding-window DFT and numerically computes its time derivative $dV_\mathrm{SCV1}/dt$ using the method of central finite difference. A PSB signal is issued based on a main PSB function that is only allowed to operate after first receiving an activation signal. An OST signal is issued based on the number of zero crossings of $V_\mathrm{SCV1}$.

\subsection{Main PSB Function for Continuous Swing Cycles}
\label{sec:Proposed-PSB-cont}
For continuous blocking, a PSB signal is issued if
\begin{equation}
    \Bigg|\frac{dV_\mathrm{SCV1}}{dt}\Bigg|\geq\Bigg|-\frac{E}{2}\sin(\delta_\mathrm{load,max}/2)\omega_\mathrm{slip,min}\Bigg|,
    \label{PSB-cont}
\end{equation}
where $\delta_\mathrm{load,max}$ is the load angle required to transfer the nominal power over the line, $\omega_\mathrm{slip,min}$ is the slowest power swing considered to be blocked by the PSB and $E=1$~pu, making tuning easy based on pre-fault normal operating conditions. First, $\delta_\mathrm{load,max}$ can be obtained from pre-fault load flow analysis. Second, $\omega_\mathrm{slip,min}=2\pi f_\mathrm{slip,min}$ can be a low value corresponding to extremely slow power swings where $\bar{Z}_1$ may enter the distance zones, such as $f_\mathrm{slip}\in(0.1,0.2)$~Hz, but also to pickup if $|dV_\mathrm{SCV1}/dt|$ stalls due to IBRs during a swing, such as in Fig.~\ref{fig:SCV-approx}~(b). Note that (\ref{PSB-cont}) does not enforce any upper limit on the swing frequency or movement speed of $\bar{Z}_1$, which is beneficial concerning IBRs and fast swings.

A minimum $\omega_\mathrm{slip}$ may be 1~rad/s ($f_\mathrm{slip,min} = 0.16$~Hz), which was used in Fig.~\ref{fig:SCV-approx}. If (\ref{SCV-approx-eq}) accurately reflects (\ref{VSCV1}), as in Fig.~\ref{fig:SCV-approx}~(a), (\ref{PSB-cont}) activates for $\delta_\mathrm{load,max}<\delta<2\pi-\delta_\mathrm{load,max}$ when the swing frequency is equal to $f_\mathrm{slip,min}$. Hence, it is only when $\bar{Z}_1$ is in the load region during $f_\mathrm{slip,min}$ power swings when (\ref{PSB-cont}) will not issue a PSB signal, where distance relay blocking is not necessary. For faster swing dynamics, the curves of $|dV_\mathrm{SCV1}/dt|$ and $|d\hat{V}_\mathrm{SCV1}/dt|$ in Fig.~\ref{fig:SCV-approx} become lifted to higher values, meaning that a PSB signal will be issued for $\delta$ lower than $\delta_\mathrm{load,max}$ and higher than $2\pi-\delta_\mathrm{load,max}$. Because it is not necessary to block the distance relay unless $\bar{Z}_1$ is inside a distance zone, (\ref{PSB-cont}) is set to only be issued if $\bar{Z}_1$ is inside a supervisory zone covering them.

\subsection{PSB Activation Signal With Swing-in-Progress Logic}
\label{sec:Proposed-PSB-Activation}
The main PSB function (\ref{PSB-cont}) is susceptible to causing unwanted blocking of distance relays during the sub-transient fault inception period and if $\delta$ accelerates during faults. The latter case is especially problematic for high-resistance faults because $\bar{Z}_1$ may move around significantly in the impedance plane, implying the power factor $\cos(\varphi_\mathrm{s})$ is varying and producing a $|dV_\mathrm{SCV1}/dt|$ possibly surpassing the low threshold corresponding to $\delta_\mathrm{load,max}$ and $\omega_\mathrm{slip,min}$ in (\ref{PSB-cont}). To prevent this yet also achieve fast PSB during the $\bar{Z}_1$ fault-to-swing transition period and first swing cycle, a PSB activation signal is implemented. 

First, the PSB scheme looks for a value of $|dV_\mathrm{SCV1}/dt|>5$~pu/s and stores the moment in time as $t_\mathrm{f}$, which approximates the fault inception time. Second, (\ref{PSB-cont}) is activated at time $t_\mathrm{PSB}$ based on one of the following criteria: 
\begin{itemize}
    \item If $\text{sgn}\big(V_\mathrm{SCV1}(t_\mathrm{f}-2T)\big) = 1$ (initially a load-out condition), activate (\ref{PSB-cont}) when $dV_\mathrm{SCV1}/dt > 5$~pu/s
    \item If $\text{sgn}\big(V_\mathrm{SCV1}(t_\mathrm{f}-2T)\big) = -1$ (initially a load-in condition), activate (\ref{PSB-cont}) when $dV_\mathrm{SCV1}/dt < -5$~pu/s
\end{itemize}
$t_\mathrm{PSB}$ estimates the fault clearance time. The location of $\bar{Z}_1$ does not matter for the PSB activation. However, once activated, a swing-in-progress logic maintains the PSB signal if $\bar{Z}_1$ is inside the supervisory zone of (\ref{PSB-cont}) at $t_\mathrm{PSB}$. This allows blocking of distance zone 1 in case $\bar{Z}_1$ immediately enters it during its fault-to-swing transition period when the sign of $dV_\mathrm{SCV1}/dt$ might change, thereby crossing zero and shortly making (\ref{PSB-cont}) not being fulfilled. 

\begin{figure}[h]
    \centering
    \includegraphics[width=0.8\linewidth]{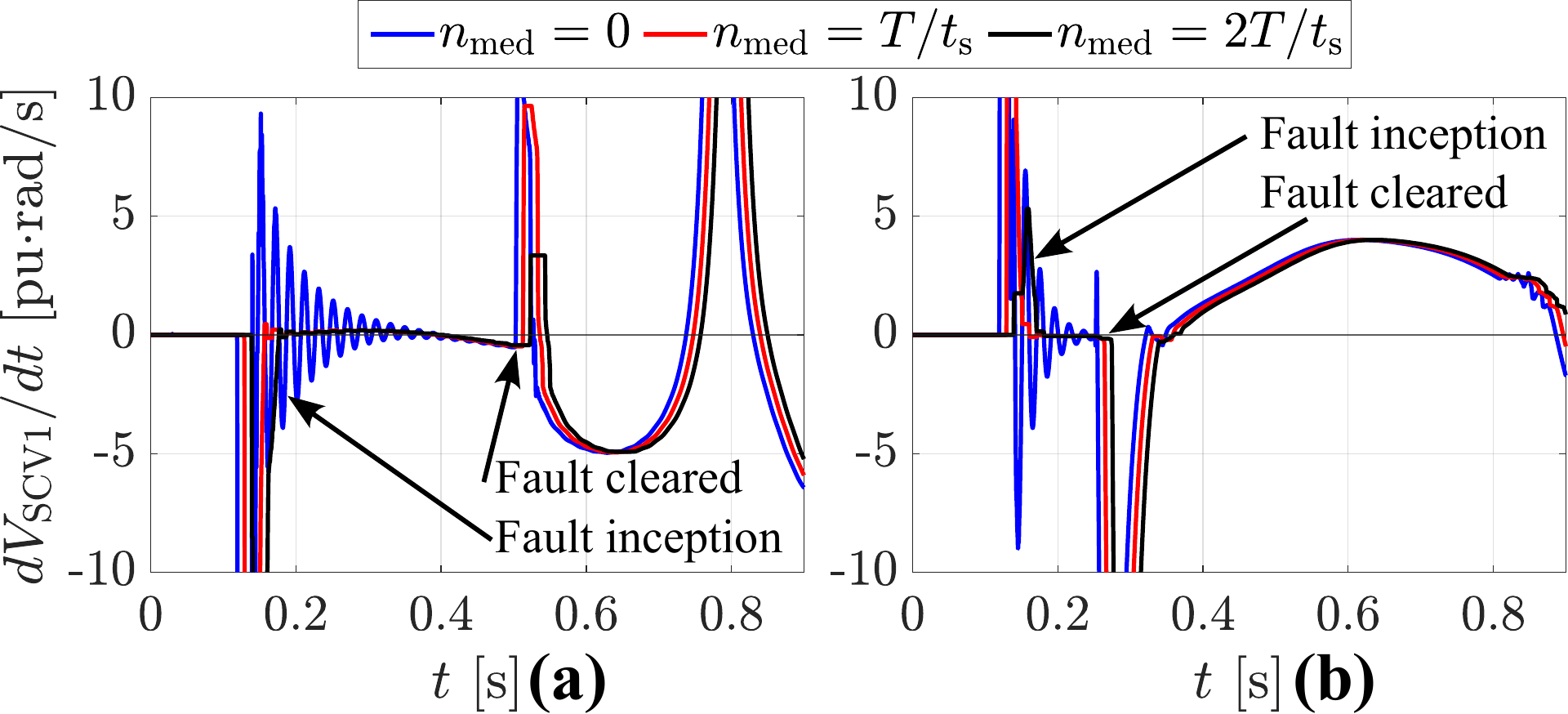}
    \caption{Impact of median filter on $dV_\mathrm{SCV1}/dt$: \textbf{(a)} load-out; \textbf{(b)} load-in.}
    \label{fig:Med-examples}
\end{figure}

For the PSB activation function, it is not reasonable to treat $dV_\mathrm{SCV1}/dt$ as a swing---it is rather the speed at which $V_\mathrm{SCV1}$ transitions from the value it had right before the fault clearance to its initial value corresponding to $\delta$ when the fault is cleared. This transition takes about 20~ms (one fundamental period $T$) due to the DFT. Hence, if $V_\mathrm{SCV1}$ goes from 0 to 1~pu over 20~ms, its rate of change becomes $1/0.02=50$~pu/s (this value is rarely surpassed at the fault clearance). It is desired to have a low value on the activation function threshold in order to pick up when $V_\mathrm{SCV1}$ does not change much upon fault clearance. A value of $5$~pu/s corresponds to only a $0.1$~pu amplitude change over 20~ms when the fault is cleared. However, using 5~pu/s risks that the fault inception triggers the PSB activation function. To mitigate this problem, a sliding-median filter of order $n_\mathrm{med}$ is applied to $dV_\mathrm{SCV1}/dt$. Fig.~\ref{fig:Med-examples} shows its effectiveness for different filter lengths, where $t_\mathrm{s}$ is the sampling period. Using $n_\mathrm{med}=T/t_\mathrm{s}$ effectively prevents unwanted PSB activation during faults and allows for fast activation when the fault is cleared (around $T/2=10$~ms) without suppressing $dV_\mathrm{SCV1}/dt$ to any significant extent. With the median filter, no implemented time delay is required between $t_\mathrm{f}$ and the activation signal at time $t_\mathrm{PSB}$.

The proposed SCV-based PSB activation function triggers with ease for the majority of power system fault-to-swing transitions listed in Section~\ref{sec:benchmark-performance-eval}. However, very long-lasting faults can cause $V_\mathrm{SCV1}$ to not change much or with opposite signs compared to the expected load-in and load-out conditions upon fault clearance. To still reliably and quickly activate the PSB function for such cases, a backup PSB activation is implemented based on a similar logic as proposed here but using higher-order incremental currents \cite{HJ-PSB-IQ2}. They remain near zero during long-lasting faults and experience sharp transients upon fault clearance. This PSB activation function is not used as the main one because it requires an inherent delay of several cycles before it has any chance of operating \cite{HJ-PSB-IQ2}.

\subsection{Proposed OST Function}
\label{sec:Proposed-OST}
The proposed OST function is based on counting the number of zero crossings $n_0$ of $V_\mathrm{SCV1}$ after time $t_\mathrm{PSB}$, regardless of where $\bar{Z}_1$ is in the impedance plane. Having no requirements on the $\bar{Z}_1$ location is beneficial because $\bar{Z}_1$ can move in trajectories that do not come near the blinders or supervisory zones, particularly in IBR systems where multiple lines and sources exist behind the relay \cite{PSB-Yongxin-GFM, PSB-Yongxin-GFL}. For such unstable swing conditions, relays equipped with OST should pick up to execute a controlled separation of the power system. 

From the analysis in Section~\ref{sec:Impact-IBRs-SCV}, $V_\mathrm{SCV1}$ makes two zero crossings per swing cycle with either an SG or IBR connected in Fig.~\ref{fig:two-source-swing} (one is for the $P\mathrm{s}$ reversal when $\bar{Z}_1$ wraps around), although the exact $\delta$ at which they occur varies, as shown with GFM inverters in Fig.~\ref{fig:SCV-approx} and GFL inverters in (\ref{VSCV1-GFL}). Because IBRs have the possibility of synchronizing even after $\delta$ completes a full cycle and $V_\mathrm{SCV1}$ makes two zero crossings, a minimum of $n_0=3$ zero crossings is proposed for the OST concerning IBRs. Such stable swings can occur from droop-based GFM inverters with their first-order synchronization dynamics, or GFL inverters entering the full reactive-current control mode while following the swing dynamics of the PCC voltage caused by the sources oscillating at the other line end. Complex multi-modal power swings can cause $d\delta/dt$ to change direction during a swing cycle and cause more than two zero crossings of $V_\mathrm{SCV1}$ before $\delta$ completes a full cycle if $\bar{Z}_1$ oscillates back and forth around the reactive axis. If such scenarios are expected from stable power swings, $n_0$ could be increased, but $n_0 = 3$ is proposed as the IBR baseline. 

Counting of $n_0$ starts immediately when the PSB activation signal is triggered. However, during high-resistance faults, $V_\mathrm{SCV1}$ can already have crossed zero and make another zero crossing immediately after $t_\mathrm{PSB}$ for either load-in or load-out conditions. To prevent this zero crossing from counting towards $n_0$, the relay checks whether $\text{sgn}\big(V_\mathrm{SCV1}(t_\mathrm{f}-2T)\big)$ and $\text{sgn}\big(V_\mathrm{SCV1}(t_\mathrm{PSB})\big)$ match. If they do not, the number of $n_0$ required to issue the OST command is increased by one (from $3$ to $4$). If they match, no adjustments are made.

\section{Performance Evaluation}
\label{sec:benchmark-performance-evaluation}
The PSCAD simulations described in Section~\ref{sec:benchmark-performance-eval} are used for benchmark performance evaluation of the proposed SCV-based PSB/OST scheme, and to validate the analytical findings in Section~\ref{sec:Impact-IBRs-SCV} with IBRs. The proposed SCV-based PSB/OST scheme operates correctly for 100\% of the simulation cases. It prevents the distance relay from tripping during power swings but never during the faults, and detects all OOS conditions without misinterpreting any stable swing as unstable. 

\begin{figure}[h]
    \centering
    \includegraphics[width=0.8\linewidth]{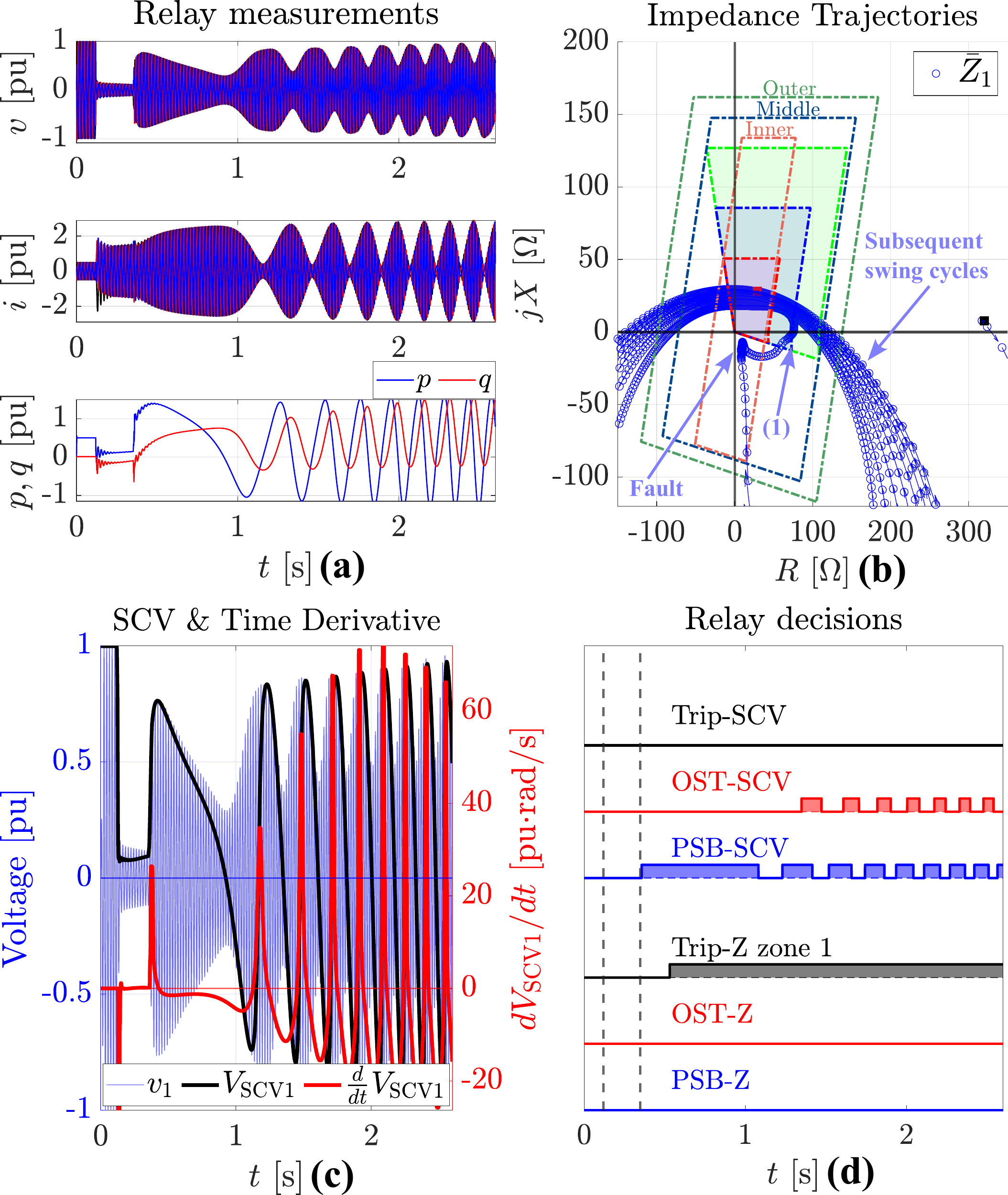}
    \caption{Unstable swing from an SG observed in Rel1: \textbf{(a)} relay measurements; \textbf{(b)} $\bar{Z}_1$; \textbf{(c)} SCV; \textbf{(d)} operation of PSB/OST schemes and distance zones.}
    \label{fig:ABCG_10_0_0_1_0.25_3_1SG}
\end{figure}

Fig.~\ref{fig:ABCG_10_0_0_1_0.25_3_1SG} shows the performance of the blinder- and SCV-based PSB/OST schemes at Rel1 for an unstable power swing produced with an SG in Fig.~\ref{fig:two-source-parallel-ohl}. The two dashed lines in Fig.~\ref{fig:ABCG_10_0_0_1_0.25_3_1SG}~(d) mark the fault inception and clearance, respectively. As can be seen in Fig.~\ref{fig:ABCG_10_0_0_1_0.25_3_1SG}~(b), $\bar{Z}_1$ never leaves the blinders during the first oscillating cycle, causing distance zone 1 to trip with the blinder-based PSB scheme. However, the proposed SCV-based PSB scheme correctly prevents the unwanted tripping. Moreover, with $\Delta T_\mathrm{PSB,1}=30$~ms, the blinders never pick up on the subsequent swing cycles, leading to the OST not activating. However, the SCV-based OST picks up correctly. 

\begin{figure}[h]
    \centering
    \includegraphics[width=0.83\linewidth]{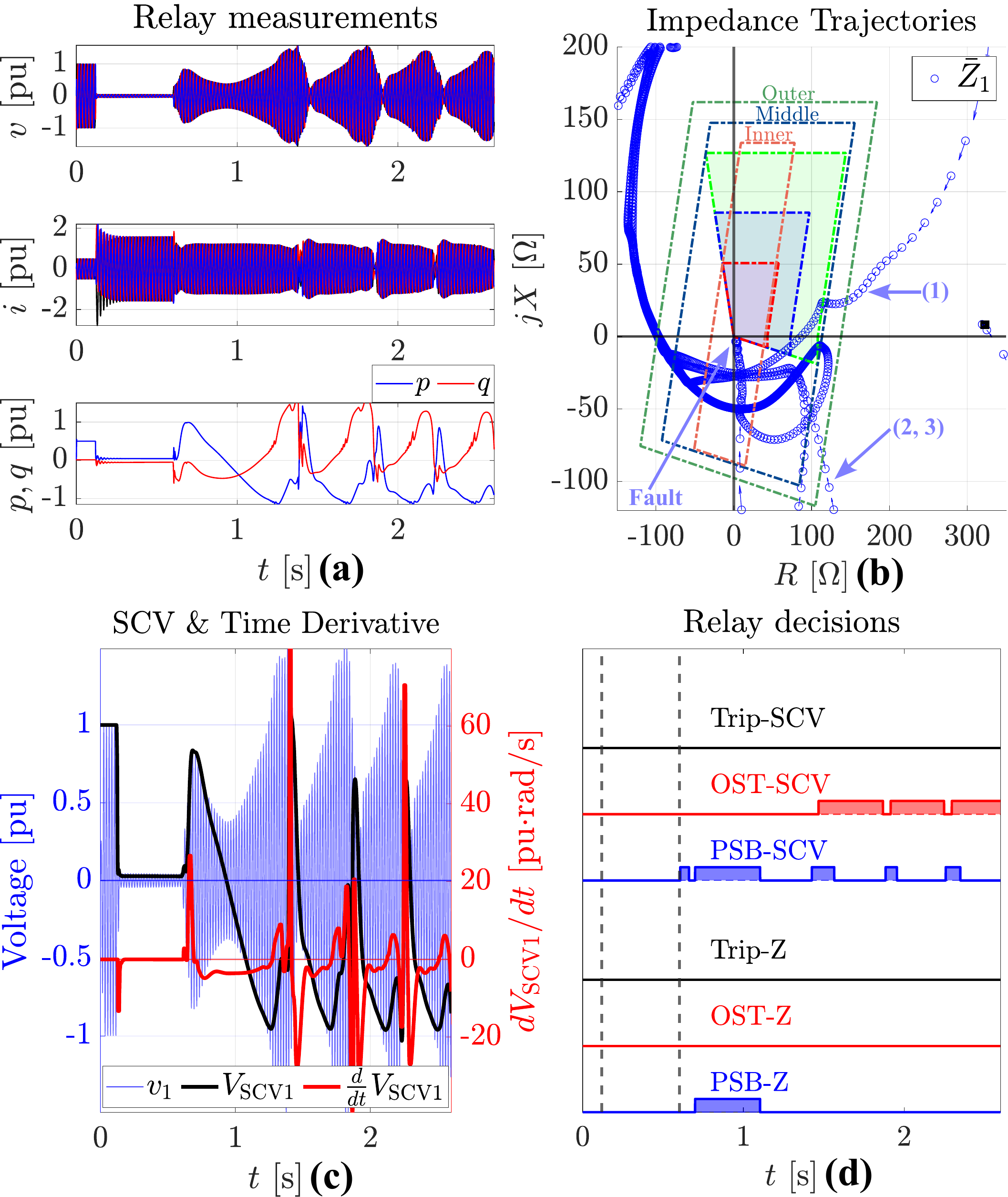}
    \caption{Unstable swing from a VSG-based GFM inverter observed in Rel1.}
    \label{fig:ABCG_10_0_0_1_0.5_VSG_1GFM1}
\end{figure}

During unstable swings produced from the VSG-based GFM inverter, the proposed SCV-based OST scheme picks up when the blinder-based OST does not, for example as shown in Fig.~\ref{fig:ABCG_10_0_0_1_0.5_VSG_1GFM1}. The implemented GFM inverter has higher stability margins than the SG due to the active power reference being reduced during faults; however, the 500~ms close-in fault made it go unstable. It can be observed in Fig.~\ref{fig:ABCG_10_0_0_1_0.5_VSG_1GFM1} that the swing frequency is not high---it is indeed slower than in Fig.~\ref{fig:ABCG_10_0_0_1_0.25_3_1SG} with the SG. However, shortly after the high positive spikes in $dV_\mathrm{SCV1}/dt$ caused by $P_\mathrm{s}$ reversal from negative to positive, $dV_\mathrm{SCV1}/dt$ accelerates to high negative values because the current limiter gets engaged, matching the high negative spike around $\delta = 45^\circ$ in Fig.~\ref{fig:SCV-approx}~(b) and premature zero crossing of $V_\mathrm{SCV1}$ due to it quickly plummeting at the start of each swing cycle. This causes $\bar{Z}_1$ to accelerate precisely when crossing the blinders for each swing cycle, agreeing with the analytical findings in (\ref{VSC1-GFM-sat}) and (\ref{Is1-GFM-sat}). It only takes 3-5~ms for $\bar{Z}_1$ to cross between the outer and middle blinders at the lower-right corner (similar speeds have been observed from field cases with IBRs \cite{PSB-CFE-SCV_vs_Blinders}), and around 9~ms when crossing in the first quadrant. Hence, the blinder-based PSB scheme does not pick up, and consequently also the OST, unlike the proposed SCV-based PSB/OST scheme. $\bar{Z}_1$ traverses on the left side of the blinders due to the EC being shifted left by the current limiter, as explained in Section~\ref{sec:Impact-IBRs-SCV}. Moreover, $dV_\mathrm{SCV1}/dt$ stalls when the current limiter virtual resistance decreases after $\bar{Z}_1$ has passed the EC, causing $dV_\mathrm{SCV1}/dt$ to get a minor positive spike followed by negative values and lastly a high positive spike caused by $P_\mathrm{s}$ reversal, all in agreement with Fig.~\ref{fig:SCV-approx}~(b) and the analysis in Section~\ref{sec:Impact-IBRs-SCV}.

\begin{figure}[h]
    \centering
    \includegraphics[width=0.83\linewidth]{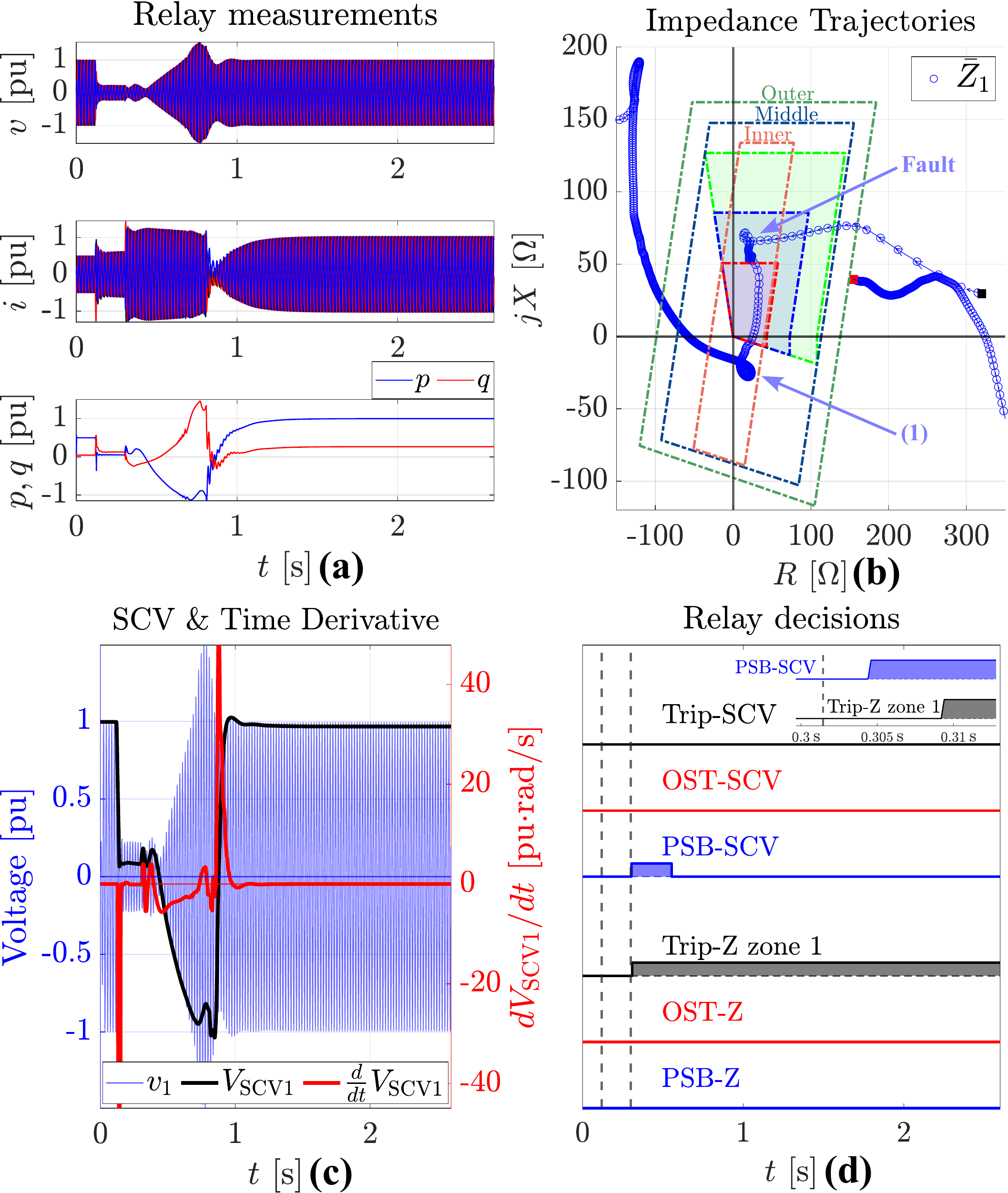}
    \caption{Stable swing from a droop-based GFM inverter observed in Rel1.}
    \label{fig:ABCG_140_0_0_1_0.2_droop_figs_SCV}
\end{figure}

Fig.~\ref{fig:ABCG_140_0_0_1_0.2_droop_figs_SCV} shows the performance of the proposed PSB/OST schemes in Rel1 for a stable power swing produced with a droop-based GFM inverter after a zone 2 fault is cleared on the parallel line in Fig.~\ref{fig:two-source-parallel-ohl}. As can be seen, the $\bar{Z}_1$ fault-to-swing transition trajectory quickly enters distance zone 1, causing it to trip if the PSB scheme is based on conventional blinders. On the other hand, the proposed PSB scheme successfully blocks distance zone 1. A zoomed-in plot of its adequate operating speed is shown in the upper right corner of Fig.~\ref{fig:ABCG_140_0_0_1_0.2_droop_figs_SCV}~(d). The SCV-based OST scheme correctly restrains. Fig.~\ref{fig:ABCG_140_0_0_1_0.2_droop_figs_SCV}~(c) shows that $V_\mathrm{SCV1}$ does not change much upon fault clearance. This is caused by the initial value of $\delta$ for the swing being such that $\text{Re}(\bar{Z}_1)\approx 0$ after the 20~ms fault-to-swing transition period with this GFM inverter emulating a high $R_\mathrm{s1}$ for current limitation, causing $V_\mathrm{SCV1}\approx 0$ both right before and after the fault clearance. However, the backup PSB activation function based on higher-order incremental currents picks up for these difficult situations \cite{HJ-PSB-IQ2}. Fig.~\ref{fig:ABCG_140_0_0_1_0.2_droop_figs_SCV}~(a) shows the first-order resynchronization dynamics of active power $p(t)$ (no overshoot, no oscillations) given by the droop-based PSC. 

\begin{figure}[h]
    \centering
    \includegraphics[width=1\linewidth]{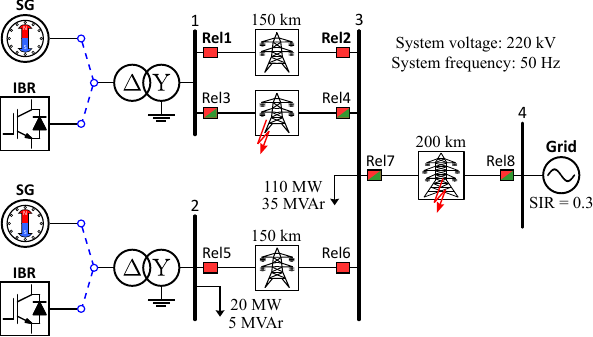}
    \caption{Three-source system to generate power swings.}
    \label{fig:three-source-system}
\end{figure}

The SCV behavior with the implemented GFL inverter connected in Fig.~\ref{fig:two-source-parallel-ohl} matches the analytical analysis in Section~\ref{sec:Impact-IBRs-SCV}, and the proposed SCV-based PSB/OST scheme overcomes limitations with the blinder-based scheme. A three-source system is also implemented in PSCAD for further performance evaluation. Fig.~\ref{fig:three-source-system} shows the system. It has a topology similar to the multi-source system used for power swing examples in the distance relay manual \cite{REL670-application-manual}. Power swings are caused by applying the balanced faults listed in Section~\ref{sec:benchmark-performance-eval} on either the line protected by Rel3/Rel4 or Rel7/Rel8, and then disconnecting the faulted line. \textbf{These fault-to-swing scenarios will be denoted F34 and F78}. $m_\mathrm{f}$ is referred to Rel3 and Rel7 for the F34 and F78 faults, respectively. The sources at bus 1 and 2 are either SGs, GFL inverters or GFM inverters (droop or VSG), or combinations. The SCV-based PSB/OST scheme is evaluated in Rel1 and Rel2, where it demonstrates superior performance to conventional blinders (for both F34 and F78).

\begin{figure}[h]
    \centering
    \includegraphics[width=0.85\linewidth]{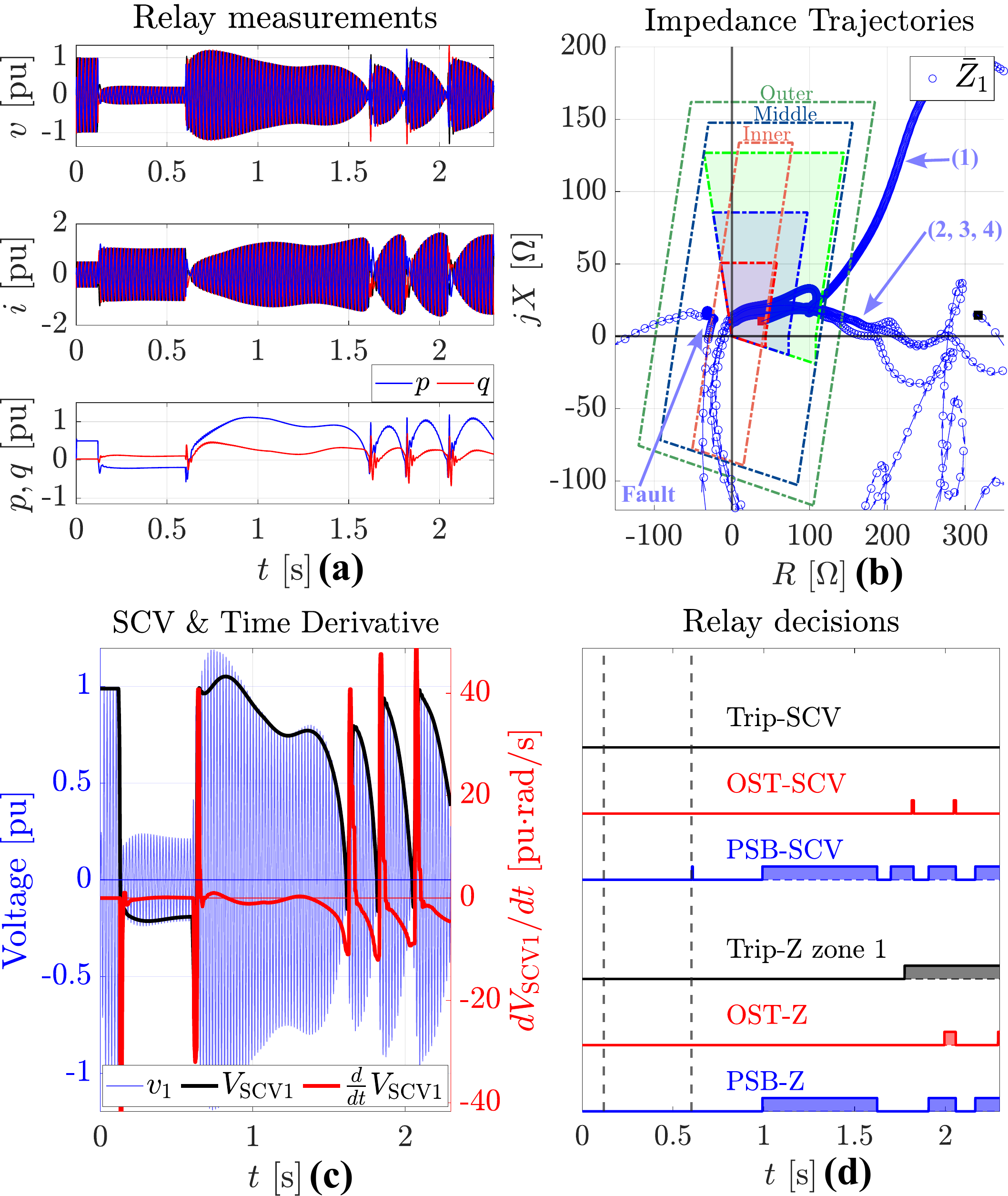}
    \caption{Rel1: GFL inverter at bus 1 and SG at bus 2, \textbf{F34} fault.}
    \label{fig:ABCG_1_10_15_0_1_0.5_H3_GFL}
\end{figure}

Fig.~\ref{fig:ABCG_1_10_15_0_1_0.5_H3_GFL} shows the performance of the proposed and blinder-based PSB/OST schemes in Rel1 after a close-in high-resistance \textbf{F34} fault is cleared when a GFL inverter is connected at bus 1 and an SG at bus 2 in Fig.~\ref{fig:three-source-system}. The blinders manage to pick up on the first, third and fourth swing cycles, but miss the second swing cycle due to $\bar{Z}_1$ crossing the dual blinders in $9~\text{ms}<\Delta T_\mathrm{PSB,2}=10$~ms, causing distance zone 1 to trip. However, the proposed SCV-based PSB/OST scheme operates correctly. It can be seen in Fig.~\ref{fig:ABCG_1_10_15_0_1_0.5_H3_GFL}~(c) that $V_\mathrm{SCV1}$ plummets to 0 and becomes negative when the GFL inverter goes into full reactive-current prioritization after the positive-sequence voltage $v_1(t)$ dips, increasing $|d\bar{Z}_1/dt|$. $V_\mathrm{SCV1}$ only becomes slightly negative because $\theta_\mathrm{PLL}$ only slightly overshoots $\delta_\mathrm{r}$ (\ref{VSCV1-GFL}) when tracking $\theta_\mathrm{s1}$. When $|\bar{V}_\mathrm{s1}|$ recovers and $P_\mathrm{s}>0$ is restored, $V_\mathrm{SCV1}$ quickly rises to positive values and gets its second zero crossing. Despite the fast nonlinear dynamics occurring in $v(t)$, $i(t)$, $p(t)$ and $q(t)$, as well as the PLL tracking an oscillating voltage caused by the SG at bus 2, $V_\mathrm{SCV1}$ has two smooth zero crossings per swing cycle. All of this is in agreement with (\ref{VSCV1-GFL}) and (\ref{theta-tilde-PLL-delta-slip}) from the SCV analysis with GFL inverters in Section~\ref{sec:Impact-IBRs-SCV}.

\section{Conclusion}
\label{sec:conclusion}
This paper analyzes the characteristics of the SCV and the impacts of GFM and GFL inverters on its local estimation in distance relays. It is analytically shown that the local estimation matches the true SCV with GFM inverters emulating an additional inductive source impedance for current limitation, but not if the current limiter emulates a high source resistance. Such current limiters accelerate the SCV and $\bar{Z}_1$ rate of change when the load angle rotates towards the $180^\circ$ antiphase condition, causing interoperability issues with blinder-based PSB/OST schemes. GFL inverters also have similar impacts when entering the full reactive-current control mode. Based on the analytical analysis, an SCV-based PSB/OST scheme is proposed, which provides fast PSB activation during the fault-to-swing transition period, remains secure during long-lasting high-resistance faults, reliably blocks the distance zones during power swings and detects OOS conditions. The main PSB criterion can easily be tuned from simple system parameters. The OST function is motivated from the analytical analysis of the SCV estimation with IBRs and requires only a single parameter for tuning. Extensive PSCAD/EMTDC simulations verify the SCV analytical analysis with GFM and GFL inverters, and the superior performance of the proposed PSB/OST scheme compared to blinders in two- and three-source systems.

\vfill

\end{document}